\documentclass[longauth]{aa}  

\usepackage{graphicx}
\usepackage{xcolor,soul}
\usepackage{svg}
\usepackage{longtable}
\usepackage{array}
\usepackage{lscape}
\usepackage{adjustbox}
\usepackage{xcolor}
\usepackage[toc,page]{appendix}

\usepackage{txfonts}
\usepackage[breaklinks, colorlinks, citecolor=blue, linkcolor=blue]{hyperref}
\usepackage{amsmath,siunitx}

\providecommand{\bjdtdb}{\ensuremath{\rm {BJD_{TDB}}}}

\providecommand{\mst}{\ensuremath{\,{\rm M_\odot}}}
\providecommand{\rst}{\ensuremath{\,{\rm R_\odot}}}

\providecommand{\arcsec}{$^{\prime \prime}$}

\begin{document}

   \title{TOI-1268b: the youngest, hot, Saturn-mass transiting exoplanet
   }

   \subtitle{}

   \author{J.\,\v{S}ubjak
          \inst{1,2,3}
          \and
          M.\,Endl
          \inst{4,5}
          \and
          P.\,Chaturvedi
          \inst{6}
          \and
          R.\,Karjalainen
          \inst{1}
          \and
          W.\,D.\,Cochran
          \inst{5,7}
          \and
          M.\,Esposito
          \inst{6}
          \and
          D.\,Gandolfi
          \inst{8}
          \and
          K.\,W.\,F.\,Lam
          \inst{9,10}
          \and
          K.\,Stassun
          \inst{11,12}
          \and
          J.\,\v{Z}\'{a}k
          \inst{3}
          \and
          N.\,Lodieu
          \inst{13,14}
          \and
          H.\,M.\,J.\,Boffin
          \inst{3}
          \and
          P.\,J.\,MacQueen
          \inst{7}
          \and
          A.\,Hatzes
          \inst{6}
          \and
          E.\,W.\,Guenther
          \inst{6}
          \and
          I.\,Georgieva
          \inst{15}
          \and
          S.\,Grziwa
          \inst{16}
          \and
          H.\,Schmerling
          \inst{16}
          \and
          M.\,Skarka
          \inst{1}
          \and
          M.\,Bla\v{z}ek
          \inst{1,17}
          \and
          M.\,Karjalainen
          \inst{1}
          \and
          M.\,\v{S}pokov\'{a}
          \inst{1,17}
          \and
          H.\,Isaacson
          \inst{18}
          \and
          A.\,W.\,Howard
          \inst{19}
          \and
          C.\,J.\,Burke
          \inst{20}
          \and
          V.\,Van Eylen
          \inst{21}
          \and
          B.\,Falk
          \inst{22}
          \and
          M.\,Fridlund
          \inst{15,23}
          \and
          E.\,Goffo
          \inst{6,8}
          \and
          J.\,M.\,Jenkins
          \inst{24}
          \and
          J.\,Korth
          \inst{25}
          \and
          J.\,J.\,Lissauer
          \inst{24}
          \and
          J.\,H.\,Livingston
          \inst{26,27,28}
          \and
          R. Luque
          \inst{29}
          \and
          A.\,Muresan
          \inst{15}
          \and
          H.\,P.\,Osborn
          \inst{20,30}
          \and
          E.\,Pall\'e
          \inst{13,14}
          \and
          C.\,M.\,Persson
          \inst{15}
          \and
          S.\,Redfield
          \inst{31}
          \and
          G.\,R.\,Ricker
          \inst{20}
          \and
          S.\,Seager
          \inst{20,32,33}
          \and
          L.\,M.\,Serrano
          \inst{8}
          \and
          A.\,M.\,S.\,Smith
          \inst{10}
          \and \\
          P. Kab\'{a}th\inst{1}}

   \institute{Astronomical Institute, Czech Academy of Sciences, Fri{\v c}ova 298, 251 65, Ond\v{r}ejov, Czech Republic
         \and
         Astronomical Institute of Charles University, V Hole\v{s}ovi\v{c}k\'ach 2, 180 00, Prague, Czech Republic
         \and
         ESO, Karl-Schwarzschild-Stra{\ss}e 2, 85748 Garching bei M\"unchen, Germany \\
         \email{jan.subjak@asu.cas.cz}
         \and
         Department of Astronomy, The University of Texas at Austin, Austin, TX, 78712, USA
         \and
         Center for Planetary Systems Habitability, The University of Texas at Austin, Austin, TX, 78712, USA
         \and
         Thueringer Landessternwarte Tautenburg, Sternwarte 5, 07778 Tautenburg, Germany
         \and
         McDonald Observatory, The University of Texas at Austin, Austin, TX, 78712, USA
         \and
         Dipartimento di Fisica, Universit\`a degli Studi di Torino, via Pietro Giuria 1, I-10125, Torino, Italy
         \and
         Centre for Astronomy and Astrophysics, Technical University Berlin, 10585 Berlin, Germany
         \and
         Institute of Planetary Research, German Aerospace Center (DLR), Rutherfordstraße 2, 12489 Berlin, Germany
         \and
         Vanderbilt University, Department of Physics \& Astronomy, 6301 Stevenson Center Ln., Nashville, TN 37235, USA
         \and
         Fisk University, Department of Physics, 1000 18th Ave. N., Nashville, TN 37208, USA
         \and
         Instituto de Astrof\'isica de Canarias (IAC), Calle V\'ia L\'actea s/n, E-38205 La Laguna, Tenerife, Spain
         \and
         Departamento de Astrof\'isica, Universidad de La Laguna (ULL), E-38205 La Laguna, Tenerife, Spain
         \and
         Department of Space, Earth and Environment, Chalmers University of Technology, Onsala Space Observatory, SE-439 92 Onsala, Sweden
         \and
         Rheinisches Institut f\"ur Umweltforschung an der Universit\"at zu K\"oln, Aachener Strasse 209, D-50931 K\"oln, Germany
         \and
         Department of Theoretical Physics and Astrophysics, Masaryk University, Kotl\'a{\v r}sk\'a 2, 61137 Brno, Czech Republic
         \and
         Astronomy Department, University of California, Berkeley, CA 94720, USA
         \and
         California Institute of Technology, Pasadena, CA 91125, USA
         \and
         Department of Physics and Kavli Institute for Astrophysics and Space Research, Massachusetts Institute of Technology, Cambridge, MA 02139, USA
         \and
         Mullard Space Science Laboratory, University College London, Holmbury St Mary, Dorking, Surrey RH5 6NT, UK
         \and
         Space Telescope Science Institute, 3700 San Martin Drive, Baltimore, MD, 21218, USA
         \and
         Leiden Observatory, Leiden University, NL-2333 CA Leiden, The Netherlands
         \and
         NASA Ames Research Center, Moffett Field, CA 94035, USA
         \and
         Department of Space, Earth and Environment, Astronomy and Plasma Physics, Chalmers University of Technology, 412 96 Gothenburg, Sweden
         \and
         Department of Astronomy, University of Tokyo, 7-3-1 Hongo, Bunkyo-ku, Tokyo 113-0033, Japan
         \and
         Astrobiology Center, 2-21-1 Osawa, Mitaka, Tokyo 181-8588, Japan
         \and
         National Astronomical Observatory of Japan, 2-21-1 Osawa, Mitaka, Tokyo 181-8588, Japan
         \and
         Instituto de Astrof\'isica de Andaluc\'ia (IAA-CSIC), Glorieta de la Astronom\'ia s/n, 18008 Granada, Spain
         \and
         NCCR/Planet-S, Universität Bern, Gesellschaftsstrasse 6, 3012 Bern, Switzerland
         \and
         Astronomy Department and Van Vleck Observatory, Wesleyan University, Middletown, CT 06459, USA
         \and
         Department of Earth, Atmospheric and Planetary Sciences, Massachusetts Institute of Technology, Cambridge, MA 02139, USA
         \and
         Department of Aeronautics and Astronautics, MIT, 77 Massachusetts Avenue, Cambridge, MA 02139, USA
             }

   \date{\today{}; \today{}}

 
\abstract{We report the discovery of TOI-1268b, a transiting Saturn-mass planet from the TESS space mission. With an age of less than one Gyr, derived from various age indicators, TOI-1268b is the youngest Saturn-mass planet known to date and contributes to the small sample of well characterised young planets. It has an orbital period of $P\,=\,8.1577080\pm0.0000044$\,days, and transits an early K dwarf star with a mass of $M_\star$\,=\,$ 0.96 \pm 0.04$\,$M_{\odot}$, a radius of $R_\star$\,=\,$ 0.92 \pm 0.06$\,$R_{\odot}$, an effective temperature of $T_\mathrm{eff}\,=\,5300\pm100$\,K, and a metallicity of $0.36\pm0.06$\,dex. By combining TESS photometry with high-resolution spectra acquired with the Tull spectrograph at McDonald observatory, and the high-resolution spectrographs at Tautenburg and Ondrejov observatories, we measured a planetary mass of $M_\mathrm{p}\,=\,96.4 \pm 8.3\,M_{\oplus}$ and a radius of $R_\mathrm{p}\,=\,9.1 \pm 0.6\,R_{\oplus}$. TOI-1268 is an ideal system to study the role of star-planet tidal interactions for non-inflated Saturn-mass planets. We used system parameters derived in this paper to constrain the planet tidal quality factor to the range of $10^{4.5-5.3}$. When compared with the sample of other non-inflated Saturn-mass planets, TOI-1268b is one of the best candidates for transmission spectroscopy studies.
}

   \keywords{planetary systems --
                spectroscopy --
                radial velocity --
                photometry --
                stellar ages
               }
\maketitle

%
%
%

\section{Introduction}\label{sec:introduction}

After the initial discovery phase, the focus of exoplanet research is now shifting to the detailed studies of the formation and evolution of planets and their atmospheres. Transiting close-in giant planets are a key to this research because it is easier to characterise them compared to smaller planets orbiting at large distances from their host stars. One such process affecting the evolution of planetary atmospheres is atmospheric erosion. \citet{Haswell12}, \citet{Staab17}, and others have shown that substantial atmospheric erosion is ongoing in a large fraction of exoplanets. 

Planetary atmospheres can be eroded via hydrodynamic escape caused by the X-ray+EUV (XUV) radiation of the host star. As summarised by \citet{Perryman18}, the hydrodynamic escape rate scales with the flux of the XUV-radiation that the planet receives. Since the XUV-flux of young stars is orders of magnitude larger than for older ones, the main erosion phase happens in the first 300--500\,Myrs for planets orbiting solar-like stars. Because of the loss of angular momentum, mainly by stellar wind, the rotation rate, and thus the activity level and its XUV-flux, declines with age \citep{Tu15}. Planets around stars younger than about 1 Gyr are ideal targets for studying the erosion of planetary atmospheres. Gas giants with a relatively low mass but a relatively large radius are particularly interesting because the erosion rate scales with the planet's surface gravity. However, only six of them orbit stars younger than 1 Gyr: Kelt-9 \citep{Gaudi17}, Kelt-17 \citep{Zhou16}, WASP-178 \citep{Rodriguez20}, Mascara-4 \citep{Dorval20}, AU Mic \citep{Plavcan20,Martioli21} and V1298\,Tau \citep{Suarez21,Poppenhaeger21}.


Giant planets are believed to form via core accretion in a protoplanetary disk at distances greater than 0.5\,au from the host stars \citep{Wuchterl00}. Such scales provide an environment with enough solid materials and gas in order to core become sufficiently massive to accrete gas and ends up as a giant planet. The giant planet may then migrate inward according to the initial conditions \citep{Coleman:2017}. During such migration, the star-planet tidal interaction plays a role in the further evolution of these gas giants, making their orbits circularised and synchronised with the host star's rotation period \citep{Hut80, Rasio96, Pont09}. The timescales of these processes can help understand the formation and evolution path of individual systems \citep{Weiss17, Persson19}. However, this is strongly limited by uncertainties of tidal quality factors for planets and stars, which are complicated to measure. This problem was discussed in \citet{Subjak20}, who were not able to precisely assess how the system was formed because of the difficulty in measuring tidal interactions. Yet, systems that are too young to be circularised and synchronised can be used to study tidal interactions and put constraints on the tidal quality factors.

Finally, close-in gas giant planets with large radii but relatively small masses that orbit bright stars are also ideal targets for atmospheric studies. The atmospheric signature of a planet is easier to detect if it has a large scale height, which depends on the temperature and surface gravity of the planet. Such planets are ideal targets for the ESA atmospheric characterisation mission ARIEL \citep[Atmospheric Remote-sensing Infrared Exoplanet Large-survey;][]{Tinetti16,Tinetti18}. ARIEL will observe 1000 preselected transiting planets, of which 50-100 will be studied intensively. The best targets for ARIEL observations are planets that are relatively warm and orbit relatively bright stars.

Here, we report a new result from the KESPRINT consortium \citep[e.g.,][]{VanEylen21,Luque21,Subjak20,Fridlund20,Persson19}, the discovery of TOI-1268b, a Saturn-mass planet orbiting a young early K dwarf star, which is an ideal target to study the atmospheric erosion and tidal interactions.


%
%
\section{Observations}
\subsection{TESS photometry}\label{sec:tess_photometry}

TESS observed TOI-1268 as part of the four Sectors 15, 21, 22, and 41. All observations were performed with the two-minute cadence mode. TESS will further observe TOI-1268 in Sectors 48 and 49. The publicly available data for TOI-1268 can be found in the Mikulski Archive for Space Telescopes (MAST)\footnote{\url{https://mast.stsci.edu/portal/Mashup/Clients/Mast/Portal.html}}, and are provided by the TESS Science Processing Operations Center (SPOC). Transit signature of TOI-1268b was detected by both the SPOC \citep{Jenkins16} and QLP \citep{Huang20,Huang2020} pipelines and alerted by the TESS Science Office on Oct 17, 2019 \citep{Guerrero21}.

\begin{figure}[!ht]
\centering
\includegraphics[width=0.48\textwidth, trim= {0.0cm 0.0cm 0.0cm 0.0cm}]{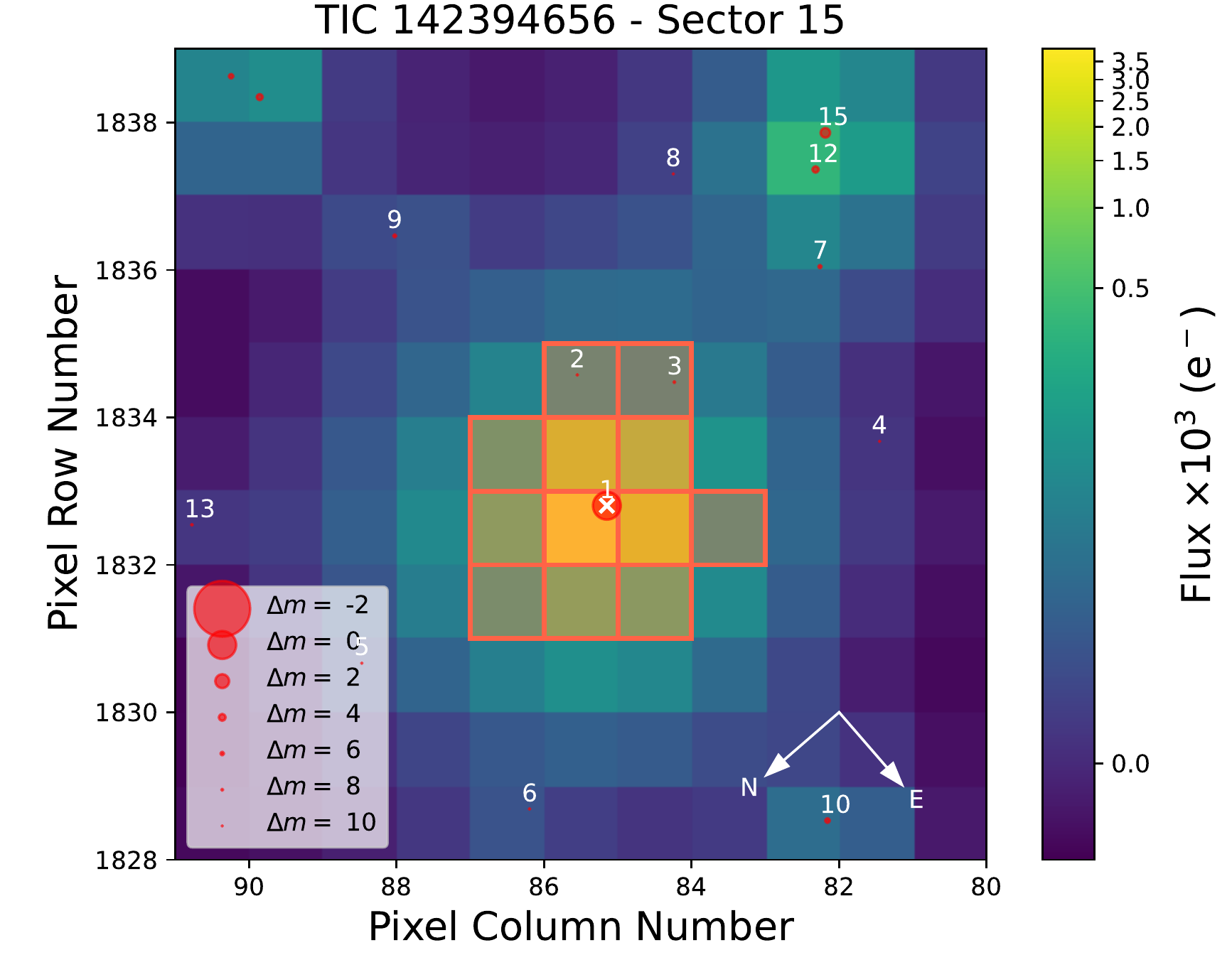}
\caption{GaiaDR2 catalog overplotted to the TESS TPF image.} \label{fig:tpf}
\end{figure}

We used the {\tt lightkurve} package \citep{Lightkurve18} to download the TESS target pixel files (Fig.~\ref{fig:tpf}) from the MAST archive directly. We then selected optimal aperture masks to obtain light curves (LCs) for each sector, which we normalised and corrected for outliers. We did not use the light curves processed by the SPOC pipeline \citep{Jenkins16}, which in addition removes the systematics of the spacecraft, as the algorithm removed one transit in Sector 15 and one transit in Sector 22. The missing transits were gapped due to scattered light features by Photometric Analysis (PA), which in this case appears to have been too aggressive. The SPOC pipeline also analyses the crowding using the Pixel Response Functions (PRFs) and includes a crowding correction in the PDC\_SAP flux time series. Not considering such a correction can lead to underestimating the planet's radius. However, the pipeline indicates that 0.9995 of the light in the optimal aperture is due to the target rather than other stellar sources suggesting the insignificant dilution due to the faint background stars. Additionally, the analysis of sectors 14-41 included a difference image centroiding analysis by Data Validation \citep{Twicken18} that indicated the source of the transit signature was within $0.375\pm2.500$\,arcsec of the target star.

\begin{figure*}[!ht]
\centering
\includegraphics[width=0.9\textwidth,height=1.05\textwidth]{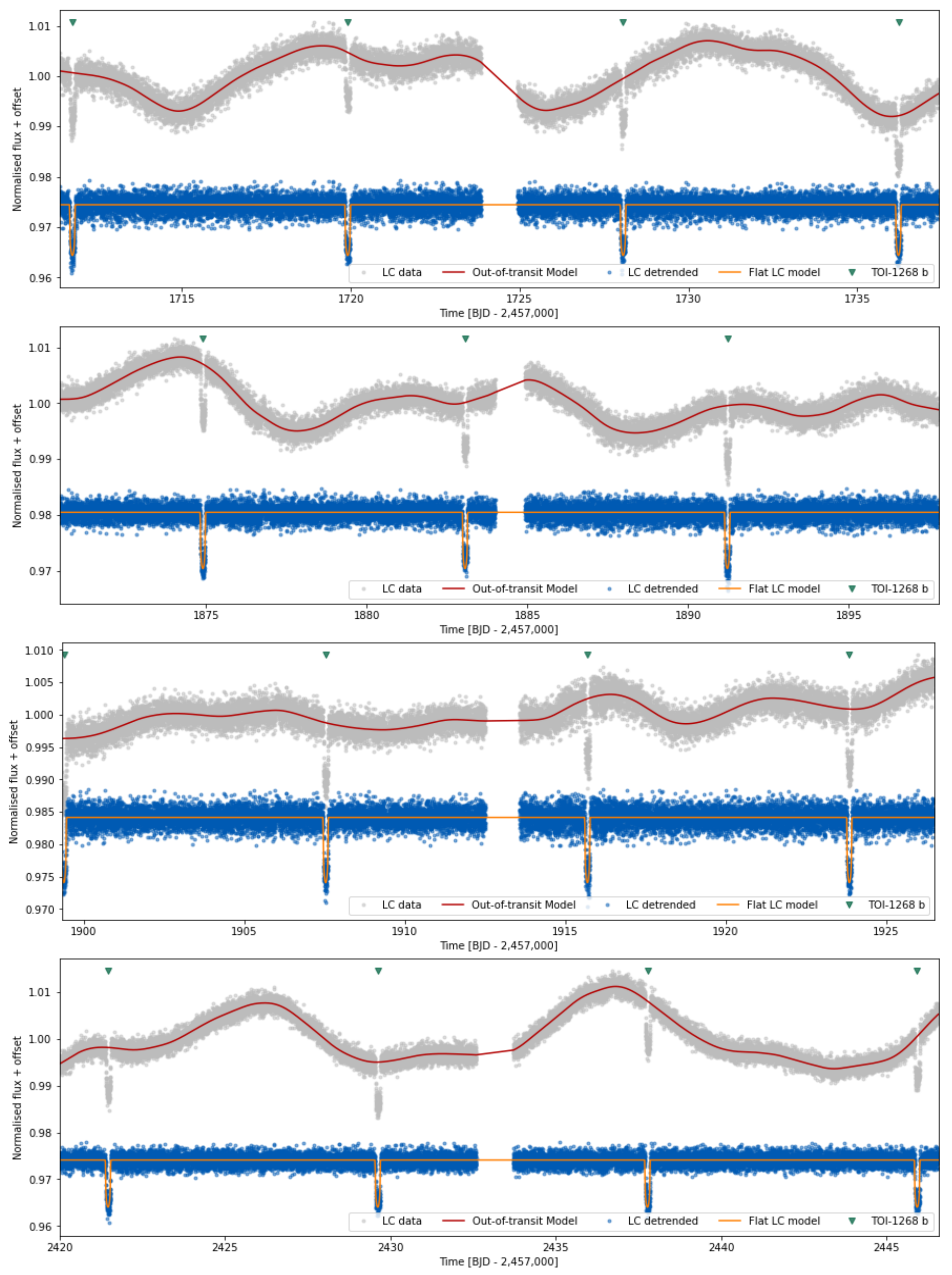}
\caption{LCs from TESS sectors for TOI-1268 created with {\tt lightkurve} from TESS tpf files. Grey points correspond to TESS observations and red lines are out-of-transit GP model created with {\tt citlalicue} following the variability in LCs. This model was substracted leading to flattened TESS LCs (blue points) with transit model (orange lines). Green triangles show the positions of transits.} 
\label{fig:tess_lc}
\end{figure*}

To correct for the systematics and remove stellar variability, we used the python package {\tt citlalicue} \citep{Barragan22} to detrend the normalised LCs extracted with {\tt lightkurve}. {\tt citlalicue} uses a Gaussian Process regression as well as transit models computed with the {\tt pytransit} code \citep{Parviainen2015} to generate a model that contains both the variability in the light curve and the transits. The variability is then removed to leave a flattened light curve with only the transit photometric variations. In this case, the variability removed contains both stellar activity and systematics. The light curves before and after the procedure are shown in Fig. \ref{fig:tess_lc}. Together 15 transits were detected, four each in Sectors 15, 22, 41, and three in Sector 21.

Additionally, we used {\tt tpfplotter} \citep{Aller20} to overplot the $Gaia$ DR2 catalogue to the TESS target pixel file (tpf) in order to identify any possible diluting sources in the TESS photometry, up to a limiting magnitude difference of 10. The tpf image created with {\tt tpfplotter} can be seen in Fig.\,\ref{fig:tpf}. We identified two additional sources between TESS pixels diluting TESS LCs. These stars are listed in Table \ref{table:sources}. With more than eight magnitude difference, these sources are too faint compared to TOI-1268 to yield any significant dilution.  The basic parameters of the star are listed in Table \ref{table:system_par}.

\begin{table}
 \centering
 \caption[]{Additional sources within the TESS aperture.
 }
 \label{table:sources}
 \begin{tabular}{@{\hspace{0mm}}l c c c@{\hspace{0mm}}}
 \hline
 \hline
$Gaia$ ID & $Gaia$ $G$ mag  & Spectral type \cr
 \hline
1675922970775714944 & 19.6 & K4--5 \cr
1675922975071294720 & 18.8 & K6--8 \cr
 \hline
 \hline

 \end{tabular}
\end{table}

\begin{table}
 \centering
 \caption[]{System parameters of TOI-1268.
 }
 \label{table:system_par}
	\begin{tabular}{lccccccr} 
		\hline
		\hline
		System       & TOI-1268 & Source\\
		\hline
        RA$_{J2000}$ (hh:mm:ss.ss) & 13 13 33.41 & 2\\
        Dec$_{J2000}$ (d:':") & 62 18 19.61 & 2\\
        \smallskip\\
        TESS $T$ mag & $10.150 \pm 0.006$ & 3\\
        $Gaia$ $G$ mag & $10.692 \pm 0.001$ & 2\\
        Tycho $B_T$ mag & $11.712 \pm 0.080$ & 4\\
        Tycho $V_T$ mag & $10.920 \pm  0.060$ & 4\\
        2MASS $J$ mag & $9.400 \pm 0.020$ & 5\\
        2MASS $H$ mag & $9.034 \pm  0.023$ & 5\\
        2MASS $K_S$ mag & $8.911 \pm 0.014$ & 5\\
        WISE1 mag & $8.886 \pm 0.023$ & 6\\
        WISE2 mag & $8.941 \pm 0.019$ & 6\\
        WISE3 mag & $8.846 \pm  0.026$ & 6\\
        WISE4 mag & $8.878 \pm 0.411$ & 6\\
        \smallskip\\
	    $\mu_\alpha\,cos(\delta)$ (mas/yr) & $-66.970 \pm 0.013$ & 1\\
	    $\mu_{\delta}$ (mas/yr) & $-15.352 \pm 0.011$ & 1\\
	    Parallax (mas) & $9.085 \pm 0.011$ & 1\\
	    U (km/s) & $-25.3 \pm 0.1$ & this work\\
	    V (km/s) & $-23.4 \pm 0.1$ & this work\\
	    W (km/s) & $10.7 \pm 0.1$ & this work\\
		\hline
		\hline
	\end{tabular}
\smallskip\\
References: 1 - $Gaia$ eDR3, with no global systematic offset applied \citep[see, e.g.,][]{StassunTorres:2021}, \citet{Gaia21}; \\ 2 - $Gaia$ DR2, \citet{Gaia18}; \\ 3 - TESS, \citet{Stassun18}; 4 - Tycho, \citet{Hog00}; 5 - 2MASS, \citet{Cutri03}; 6 - WISE, \citet{Wright10} \\
\end{table}

%
%
%

\begin{figure*}[!ht]
\centering
\includegraphics[width=1.0\textwidth]{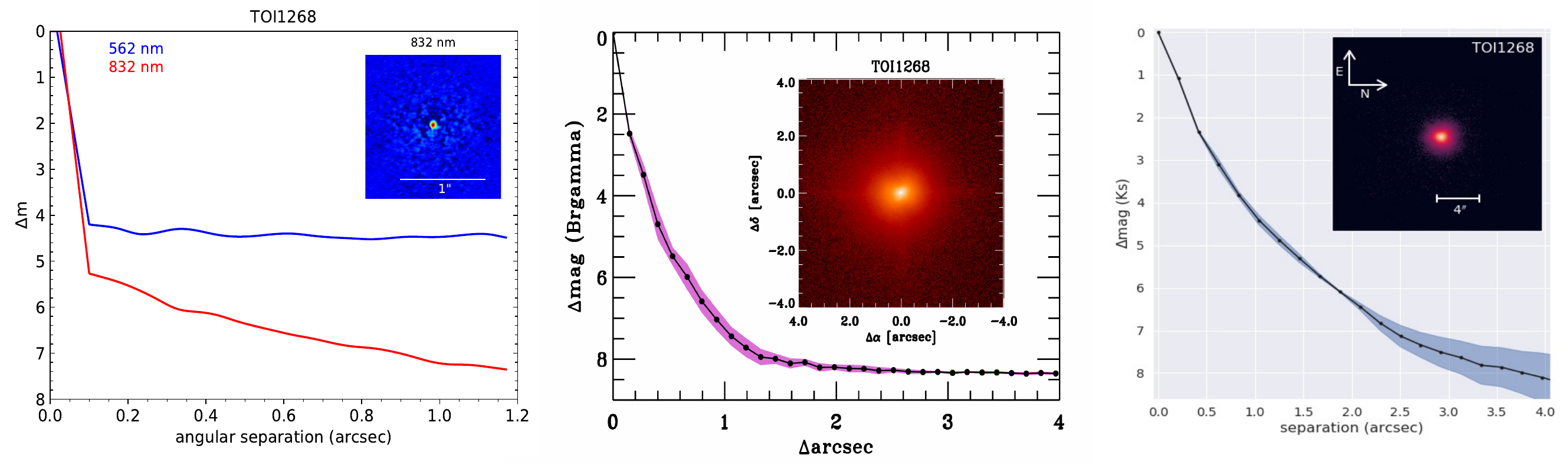}
\caption{From left to right: Alopeke contrast curve for 562\,nm and 832\,nm bands with a $1.2\arcsec \times 1.2\arcsec$ reconstructed image of the field. PHARO contrast curve for Brgamma band with a $8\arcsec \times 8\arcsec$ reconstructed image of the field. ShaneAO contrast curve for Ks band with a $20\arcsec \times 20\arcsec$ reconstructed image of the field.} \label{fig:hr_images}
\end{figure*}

\subsection{Ground-based photometry}\label{sec:ground_based_photometry}

As part of the TESS Follow-up Observing Program (TFOP), we collected ground-based photometric data of TOI-1268. The observations were scheduled using {\tt Transit Finder}, a customized version of the {\tt Tapir} software \citep{Jensen13} and photometric data were extracted using {\tt AstroImageJ} \citep{Collins17}. In a few cases, only part of the transit is observed, while in others, the light curve precision is too low to hope to improve the parameters from the TESS LCs. Hence, we do not further consider ground-based photometry in this paper.

%
%
%
%
\subsection{High-resolution imaging}\label{sec:high_resolution_imaging}

To ensure that there are no diluting sources (closer than the Gaia separation limit of 0.4\arcsec), high-resolution images were obtained, using adaptive optics and speckle imaging.

On Feb 02, 2021, TOI-1268 was observed with the Alopeke speckle imager \citep{Scott18}, mounted on the 8.1\,m Gemini-North. Alopeke uses high-speed iXon Ultra 888 back-illuminated Electron Multiplying CCDs (EMCCDs) to simultaneously acquire data in two bands centred around 562\,nm and 832\,nm. The data were reduced following the procedures in \citet{Howell11} and the final reconstructed image, shown in Fig. \ref{fig:hr_images}, reaches a contrast of $\Delta$mag = 6.36 at a separation of 0.5\arcsec in the 832\,nm band and $\Delta$mag = 4.47 at a separation of 0.5\arcsec in the 562\,nm band. The estimated PSF is 0.02\arcsec wide. At the distance of TOI-1268, the star appears single within a separation from 10 to 130\,au with contrasts between 5 to 7.5\,mag in the 832\,nm band.

On Jan 08, 2020, TOI-1268 was observed using the Palomar High Angular Resolution Observer (PHARO) \citep{Hayward01} with the JPL Palomar Adaptive Optics System, mounted on the 5.0\,m Hale telescope. PHARO uses a $1024 \times 1024$ HAWAII HgCdTe detector to observe in the 1 to 2.5\,$\mu$m range. Observations were performed with a Br$\gamma$ filter. The final reconstructed image, shown in Fig. \ref{fig:hr_images}, reaches a contrast of $\Delta$mag = 5.48 at a separation of $0.5\arcsec$ and has an estimated PSF that is 0.13\arcsec wide. The star appears single within a separation from 45 to 440\,au with contrasts between 4.5 to 8.5\,mag.

Finally, on Jan 14, 2021, TOI-1268 was observed using the SHARCS camera \citep{McGurk14} with ShaneAO, mounted on the 3.0\,m Shane telescope. ShaneAO uses $2048 \times 2048$ Teledyne HAWAII-2RG HgCdTe near-infrared detector. Observations were performed with a $Ks$ filter with a $20\arcsec$ field of view. The final reconstructed image, shown in Fig. \ref{fig:hr_images}, reaches a contrast of about $\Delta$mag = 2.65 at a separation of $0.5\arcsec$. The star appears single within a separation from 140 to 440\,au with contrasts between 4.5 to 8.5\,mag.

%
%
\subsection{Spectroscopic observations}\label{sec:spectroscopic_observations}

\subsubsection{The Tull spectrograph}

Between Dec 8, 2020 and Jul 18, 2021 we obtained a total of 32 spectra of TOI-1268 with the Tull spectrograph. The Tull cross-dispersed white-pupil spectrograph (Tull et al. 1995) is installed at the coude focus of the 2.7m Harlan J. Smith Telescope located at the McDonald Observatory. The spectrograph has a resolving power of R\,=\,60\,000 and covers wavelengths from 375\,nm to 1020\,nm. The exposure time of the observations was set to 1800\,s resulting in a signal-to-noise ratio (S/N) between 60 and 75 at 550nm, depending on the observing conditions and airmass. These spectra used an I2 vapour absorption cell as the radial velocity metric and were reduced with a pipeline script based on \textsc{IRAF} \citep{Tody86}. Radial velocities (RVs) were computed using the Austral pipeline \citep{Endl00}.

\subsubsection{The TCES spectrograph}

We obtained between Mar 4, 2020 and Jan 25, 2021 a total of 51 spectra of TOI-1268 with the Tautenburg Coud\'{e} Echelle spectrograph, attached to the 2-m Alfred Jensch telescope located of the Karl Schwarzschild Observatory. The instrument has a spectral resolving power of R\,=\,67\,000 and covers wavelengths from 467\,nm to 740\,nm. The exposure time was always set to 1800s, resulting in a typical S/N of 40 per resolution element at 550\,nm depending on the observing conditions and airmass. The spectra were calibrated with an I$_2$ vapor absorption cell and reduced with the Tautenburg Spectroscopy Pipeline – $\tau$-spline based on \textsc{IRAF} and \textsc{PyRaf} routines \citep[see][for more details]{Sabotta19}. Radial velocities were computed using the Velocity and Instrument Profile EstimatoR (VIPER\footnote{\url{https://github.com/mzechmeister/viper}}) code \citep{Zechmeister21}.

\subsubsection{The OES spectrograph}

We obtained a total of 21 spectra of TOI-1268 with the spectrograph in Ond\v{r}ejov between Aug 5, 2020 and Feb 24, 2021. The Ond\v{r}ejov Echelle Spectrograph (OES) is installed on a 2-m Perek telescope located at the Ond\v{r}ejov Observatory. The instrument has a spectral resolving power of R\,=\,50\,000 (at $500$\,nm) and covers wavelengths from 380\,nm to 900\,nm. A detailed description of the instrument can be found in \citet{Kabath20}. Exposure times were set to 3600\,s, resulting in S/N between 10--35 at 550\,nm, depending on observing conditions and airmass. Spectra were calibrated with ThAr lamp spectra acquired at the end of the night and reduced with scripts based on \textsc{IRAF}. Radial velocities were computed with the IRAF {\tt fxcor} routine.

\subsubsection{The HIRES spectrograph}

We obtained one spectrum of TOI-1268 with the HIRES spectrograph \citep{Vogt94} on Feb 23, 2021. The purpose was to get a high S/N spectrum to characterise the stellar parameters. The HIRES echelle spectrograph is installed on the 10-m Keck 1 telescope and has a spectral resolving power R\,=\,60\,000 with the C2 decker. The exposure time was 90\,s, resulting in an SNR of 45.

%
%

\section{Stellar Parameters}\label{sec:stellar_parameters}

%
%
\subsection{Stellar parameters with iSpec}\label{sec:iSpec}

We co-added all the high-resolution (R=67,000) TCES spectra taken without iodine cell and corrected for RV shifts to reach an SNR of 45 per pixel at 550\,nm. We then determined the stellar parameters of TOI-1268 by applying the Spectroscopy Made Easy radiative transfer code \citep[{\tt SME};][]{Valenti96,Piskunov17}, which is incorporated into {\tt iSpec} \citep{Blanco14,Blanco19}, on our combined spectrum. 
Complementary to it, we modelled the spectrum with MARCS models of atmospheres \citep{Gustafsson08}, which cover effective temperatures from 2500 to 8000\,K, surface gravities from 0.00 to 5.00\,dex, and metallicities from -5.00 to 1.00 dex. We also used version 5 of the GES atomic line list \citep{Heiter15}. The line list spans over the interval from 420 to 920\,nm and includes 35 chemical species. Based on these, the {\tt iSpec} then calculates synthetic spectra, which are compared to the observed one, and spectral fitting technique minimizes the $\chi^2$ value between them by executing a nonlinear least-squares (Levenberg-Marquardt) fitting algorithm \citep{Markwardt09}.

To determine an effective temperature $T_{\rm eff}$, surface gravity $\log{g}$, metallicity $\rm [Fe/H]$, and the projected stellar equatorial velocity $v\sin{i}$, we used specific features in the spectrum sensitive for these parameters. Specifically, we used the wings of H$\alpha$ line \citep{Cayrel11} to determine the effective temperature. We excluded the core of this line as it has origin in the chromosphere and hence would incorrectly result in higher temperatures. We then used the 87 Fe I,II lines between 597 and 643\,nm to determine a metallicity and projected stellar equatorial velocity. These parameters were used as inputs to the Bayesian parameter estimation code {\tt PARAM 1.3}\footnote{\url{http://stev.oapd.inaf.it/cgi-bin/param_1.3}} \citep{DaSilva06} to compute a surface gravity from PARSEC isochrones \citep{Bressan12}. The whole procedure was done several times iteratively to converge to the final values of parameters. Measuring the lithium pseudo-equivalent width $pEW_{Li}$ of the Li\,I line at 670.8\,nm as is described in Section \ref{sec:lithium} was used as an age indicator. Finally, during the modelling process in {\tt iSpec} we used empirical relations for the microturbulence and macroturbulence velocities ($V_{mic}$, $V_{mac}$) incorporated into the framework to reduce the number of free parameters. The final parameters are listed in Table \ref{table:stellar_par}.

The final parameters of $T_{\rm eff}$ and $\rm [Fe/H]$ obtained after several iterations together with the Tycho V magnitude and Gaia parallax (see Table \ref{table:system_par}) were one more time used as inputs to the {\tt PARAM 1.3} code to determine stellar mass, radius and age. To estimate TOI-1268's spectral type we used the up-to-date version\footnote{\url{https://www.pas.rochester.edu/~emamajek/EEM_dwarf_UBVIJHK_colors_Teff.txt}} of the empirical spectral type-colour sequence from \cite{Pecaut13}. 

\subsection{Stellar parameters with SpecMatch}\label{sec:SpecMatch}

\begin{figure*}[!ht]
\centering
\includegraphics[width=1.0\textwidth,height=0.6\textwidth]{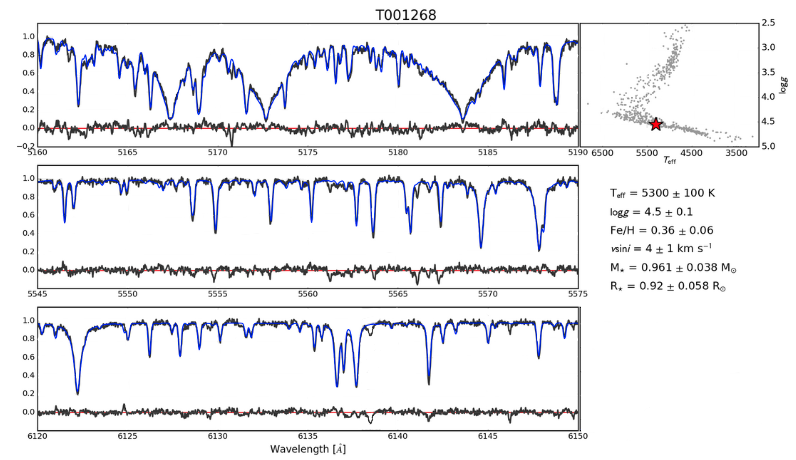}
\caption{Part of the HIRES spectrum of TOI-1268 (black) with the spectral synthesis fit (blue) and the residuals below. We also plot the position of TOI-1268 on the $logg$ vs $T_{eff}$ plane together with the {\tt SpecMatch} library of stars with high-resolution optical spectra.}
\label{fig:specmatch}
\end{figure*}

As an independent check of the stellar parameters derived above, we use also derived parameters using the HIRES spectrum and the {\tt SpecMatch} package \citep{Yee17}. To determine stellar parameters, {\tt SpecMatch} compares observed spectrum with the library of well-characterised high signal-to-noise (>\,400) HIRES spectra in combination with Dartmouth isochrones \citep{Dotter08}. All parameters are listed in Table \ref{table:stellar_par} and are in good agreement with the ones derived from {\tt iSpec}. The spectrum of TOI-1268, together with the spectral synthesis fit, is plotted in Fig. \ref{fig:specmatch}.

\begin{table}
 \centering
 \caption[]{Stellar parameters of TOI-1268.
 }
 \label{table:stellar_par}
\begin{adjustbox}{width=0.4\textwidth} 

	\begin{tabular}{lccccccr} 
		\hline
		\hline
& iSpec \& PARAM 1.3 analysis & SpecMatch \\
\hline
$\rm T_{eff}$ (K) & $5290 \pm 117$ & $5300 \pm 100$ \\
$[{\rm Fe/H}]$ (dex) & $0.34 \pm 0.11$ & $0.36 \pm 0.06$ \\
$\log{g}$ (cgs) & $4.52 \pm 0.04$ & $4.55 \pm 0.10$ \\
$v_{\rm rot} \sin{i_\star} $ (km/s) & $4.12 \pm 1.31$ & $4.12 \pm 1.00$ \\
$EW_{Li}$ (\AA) & 0.095 & \\
$M_\star$ ($\rm \mst$) & $0.92 \pm 0.03$ & $0.96 \pm 0.04$ \\
$R_\star$ ($\rm \rst$) & $0.85 \pm 0.03$ & $0.92 \pm 0.06$ \\
\hline
& VOSA analysis & \\
\hline
$\rm T_{eff} (K)$ & 5100--5300 \\
$[{\rm Fe/H}]$ & 0.0--0.5 \\
$\log{g}$ & 4--5 \\
$L_\star$ ($L_{\odot}$) & 0.50--0.52 \\
$R_\star$ ($\rm \rst$) & 0.84--0.92 \\
\hline
$P_{Rot}$ (days) & $10.9 \pm 0.5 $ \\
Spectral type & K1--K2 \\
\hline
\hline
	\end{tabular}
\end{adjustbox}
\smallskip\\
\end{table}

%
%
\subsection{SED analysis with VOSA}\label{sec:VOSA}

We modelled the Spectral Energy Distribution (SED)  using the Virtual Observatory SED Analyser \citep[{\tt VOSA}\footnote{\url{http://svo2.cab.inta-csic.es/theory/vosa/}};][]{Bayo08} as an additional independent check on the derived stellar parameters. We used grids of five different models: BT-Settl-AGSS2009 \citep{Barber06,Asplund09,Allard12}, BT-Settl-CIFIST \citep{Barber06,Caffau11,Allard12}, BT-NextGen GNS93 \citep{Grevesse93,Barber06,Allard12}, BT-NextGen AGSS2009 \citep{Barber06,Asplund09,Allard12}, and Coelho Synthetic stellar library \citep{Coelho14} to determine an effective temperature $T_{\rm eff}$, surface gravity $\log{g}$, and metallicity $\rm [Fe/H]$. We set priors for these parameters based on results from {\tt iSpec}, specifically T$_{\rm eff}$\,=\,4000--7000\,K, $\log$(g)\,=\,4.0--5.0 dex, and [Fe/H]\,=\,$-$0.5--0.5. However, priors for metallicity are limited by model used, as for example the BT-Settl-CIFIST are available only for the solar metallicity. 

\begin{figure}[!ht]
\centering
\includegraphics[width=0.52\textwidth, trim= {0.0cm 0.0cm 0.0cm 0.0cm}]{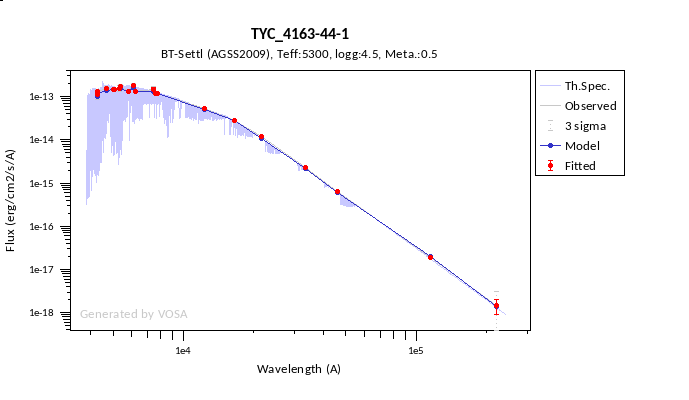}
\caption{Spectral energy distribution of TOI-1268. Red symbols represent the used photometric observations. The blue line represents the best model (BT-Settl-AGSS2009) from all different models used. Model spectrum is plotted in the background.} \label{fig:sed}
\end{figure}

We used the available photometric measurements spanning the wavelength range 0.4 – 22 $\mu$m (Fig. \ref{fig:sed}). Specifically, we used the Str{\"o}mgren-Crawford uvby{\ensuremath{\beta}} \citep{Paunzen15}, Tycho \citep{Hog00}, $Gaia$ DR2 \citep{Gaia18}, $Gaia$ eDR3 \citep{Gaia21}, 2MASS \citep{Cutri03}, AKARI \citep{Ishihara10}, and WISE \citep{Cutri14} photometry. {\tt VOSA} then uses a grid of models to compare the observed photometry with the theoretical one using ${\chi}^2$ minimization procedure. For each individual model, we take three results with the lowest ${\chi}^2$, which together create intervals of derived parameters using the lowest and highest values. We report the final intervals in Table \ref{table:stellar_par}. Additionally, {\tt VOSA} uses the effective temperature and bolometric luminosity to determine stellar radius via the Stefan–Boltzmann law. The final intervals were derived in a similar way as previously and are also reported in Table \ref{table:stellar_par}. We also used {\tt VOSA} to compare TOI-1268's SED with those in template collections provided by \citet{Kesseli17} and confirm the K1--K2 spectral type. All values derived from VOSA are in agreement with those inferred from iSpec and SpecMatch.

\begin{figure*}[!ht]
\centering
\includegraphics[width=1.0\textwidth]{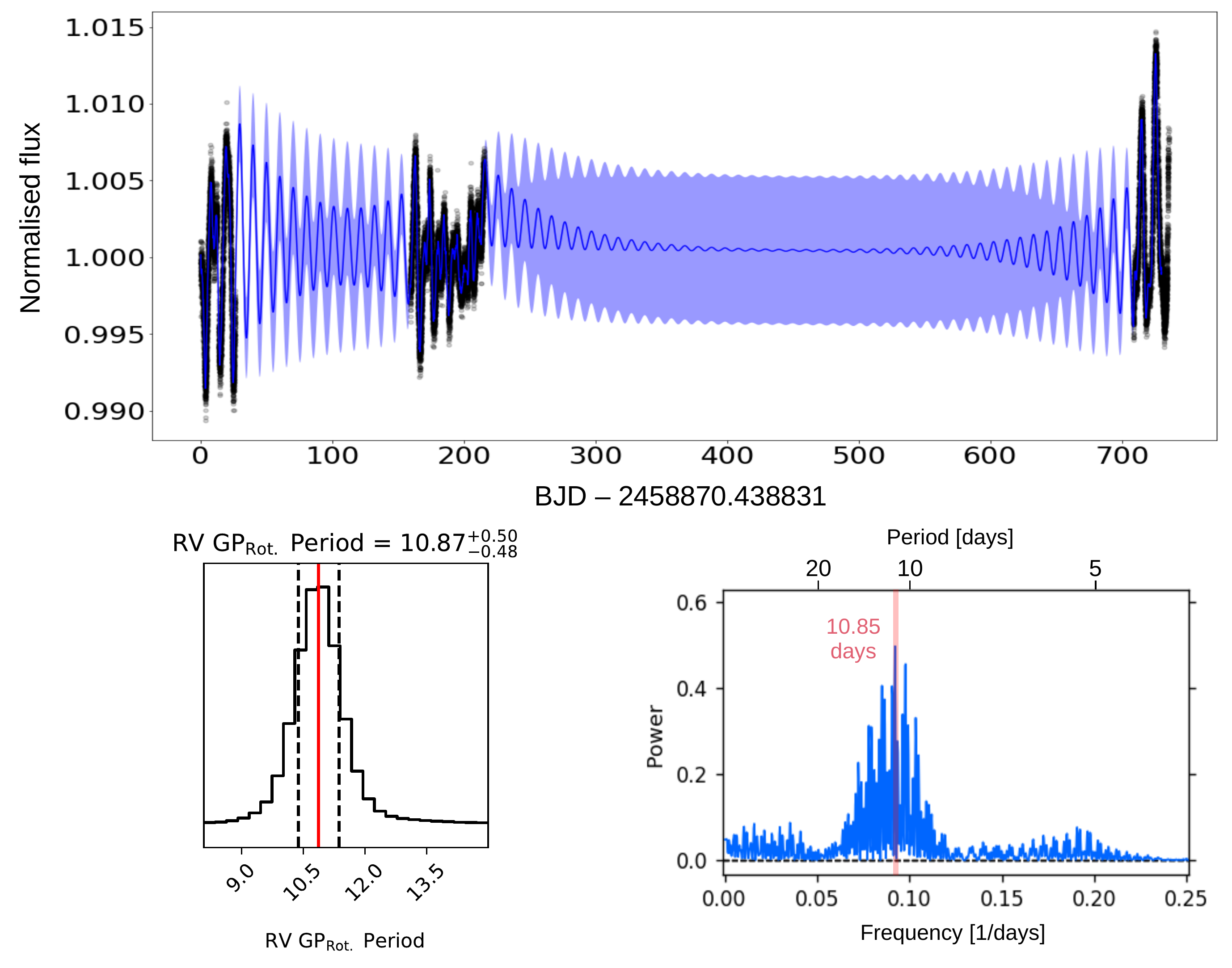}
\caption{Top: TESS data (black points) with the MAP model prediction. The blue line shows the predictive mean, and the blue contours show the predictive standard deviation. Bottom left: probability density of the rotation period. The period is the parameter $P_{rot}$ in Equation \ref{eq_rot}. The mean value is indicated by the vertical red line and 1$\sigma$ error bar is indicated with the dashed black lines. Bottom right: GLS periodogram.} \label{fig:gp_plots}
\end{figure*}

%
%
\subsection{Analysing stellar rotation}\label{sec:rotation_period}

We used the LCs derived from the TESS target pixel files (see Section \ref{sec:tess_photometry}) to determine the rotation period of the star. Before the procedure, we applied the Pixel Level Decorrelation method \citep{Deming15} to remove systematics.
We then use the Gaussian Process (GP) Regression library called {\tt Celerite}. A description of the library can be found in \cite{Foreman17b}, where the authors also discuss the physical interpretation of various kernels. To derive the rotation period through the variations in LCs caused by inhomogeneous surface features, such as spots and plages, we choose a rotational kernel function defined as:

\begin{equation}\label{eq_rot}
      k(\tau) =
         { \frac{A}{2 + B} e^{-\tau/L} \left[ cos(\frac{ 2\pi\tau }{ P_{rot} }) + (1+B)
                   \right]
}\,,
   \end{equation}

where $A$ and $B$ define the amplitude of the GP, $\tau$ is the time-lag, $L$ is a timescale for the amplitude-modulation of the GP, and $P_{rot}$ represents the rotational period. We use the L-BFGS-B nonlinear optimisation routine \citep{Byrd95,Zhu97} to estimate the maximum a posteriori (MAP) parameters. We then initialise 32 walkers and run for 1000 burn-in steps and 10000 steps of MCMC using {\tt emcee} \citep{Goodman10,Foreman13} to derive marginalised posterior distributions of free parameters. We used wide priors for A, B (log-uniform priors between $10^{-5}$ and $10^5$ ppm), L (log-uniform between $10^{-5}$ and $10^5$ days), and rotation period (uniform between 0 and 100 days). The derived rotational period from this analysis is $P_{rot}=10.9 \pm 0.5$\,days, and we plot the probability density of $P_{rot}$ together with the MAP model prediction in Fig. \ref{fig:gp_plots}. As an independent check of the derived $P_{rot}$, we also apply the generalised Lomb-Scargle (GLS) periodograms \citep{Zechmeister09} to the TESS LCs. We can see strong peaks around five and eleven days in individual sectors. Using all sectors together, the forest of peaks around eleven days is visible with the maximum at 10.9 days. We consider it to be the rotation period and its half to be the first harmonic, possibly due to spots on diametrically opposite sides of the star. We plot the periodogram in Fig. \ref{fig:gp_plots}. This value is consistent with the one inferred from the projected stellar equatorial velocity determined from spectra. For an inclination of 90 degrees, it gives a rotation period of $11.6^{+4.9}_{-2.9}$\,days. A stellar inclination close to 90 degrees would not be unexpected as for the systems where the tidal forces are expected to play a dominant role, the tidal equilibrium can be established only under assumptions of coplanarity, circularity, and synchronised rotation \citep{Hut80}. Furthermore, the scenario of orbital coplanarity is highly preferred, as we do not detect an additional object in the system. We report the derived rotation period in Table \ref{table:stellar_par}.

%
%

\section{Age analysis}\label{sec:age}

We estimate the age of TOI-1268 using several independent methods. These include stellar isochrone fitting, gyrochronology analysis, $R'_{HK}$ index, lithium equivalent width ($EW_{Li}$), and membership to young associations. Our effort is to examine each age indicator separately to provide the age interval for each of them and to investigate an overlap between these intervals.

\subsection{Stellar isochrones}

\begin{figure}[!ht]
\centering
\includegraphics[width=0.48\textwidth, trim= {0.0cm 0.0cm 0.0cm 0.0cm}]{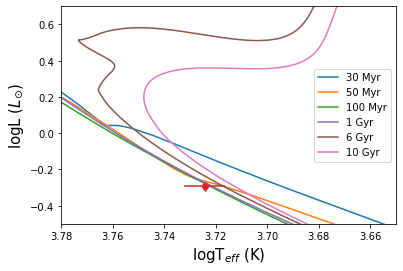}
\caption{Luminosity vs effective temperature plot. Curves represent MIST isochrones for ages: 30\,Myr (blue), 50\,Myr (orange), 100\,Myr (green), 1\,Gyr (purple), 6\,Gyr (brown), 10\,Gyr (pink), and for [Fe/H]\,=\,0.25. Red point represents the parameters of TOI-1268 with their error bars.} \label{fig:isochrones}
\end{figure}

We used the {\tt PARAM 1.3} code to derive the age of TOI-1268 based on the PARSEC isochrones. As input parameters, we used the values of $T_{\rm eff}$ and $\rm [Fe/H]$ derived in Sect. \ref{sec:iSpec}, as well as the Tycho $V$ magnitude and the Gaia-derived parallax (see Table \ref{table:system_par}). We derived an age of $3.6 \pm 3.5$\,Gyr. As an independent check, we overplot in Fig. \ref{fig:isochrones} TOI-1268's luminosity and effective temperature with the MIST stellar evolutionary tracks \citep{Choi16}. It demonstrates that we are not able to distinguish between ages from about 50\,Myr up to $\sim$6\,Gyr, as the star is on the main sequence.

\subsection{Gyrochronology}

\begin{figure}[!ht]
\centering
\includegraphics[width=0.48\textwidth, trim= {0.0cm 0.0cm 0.0cm 0.0cm}]{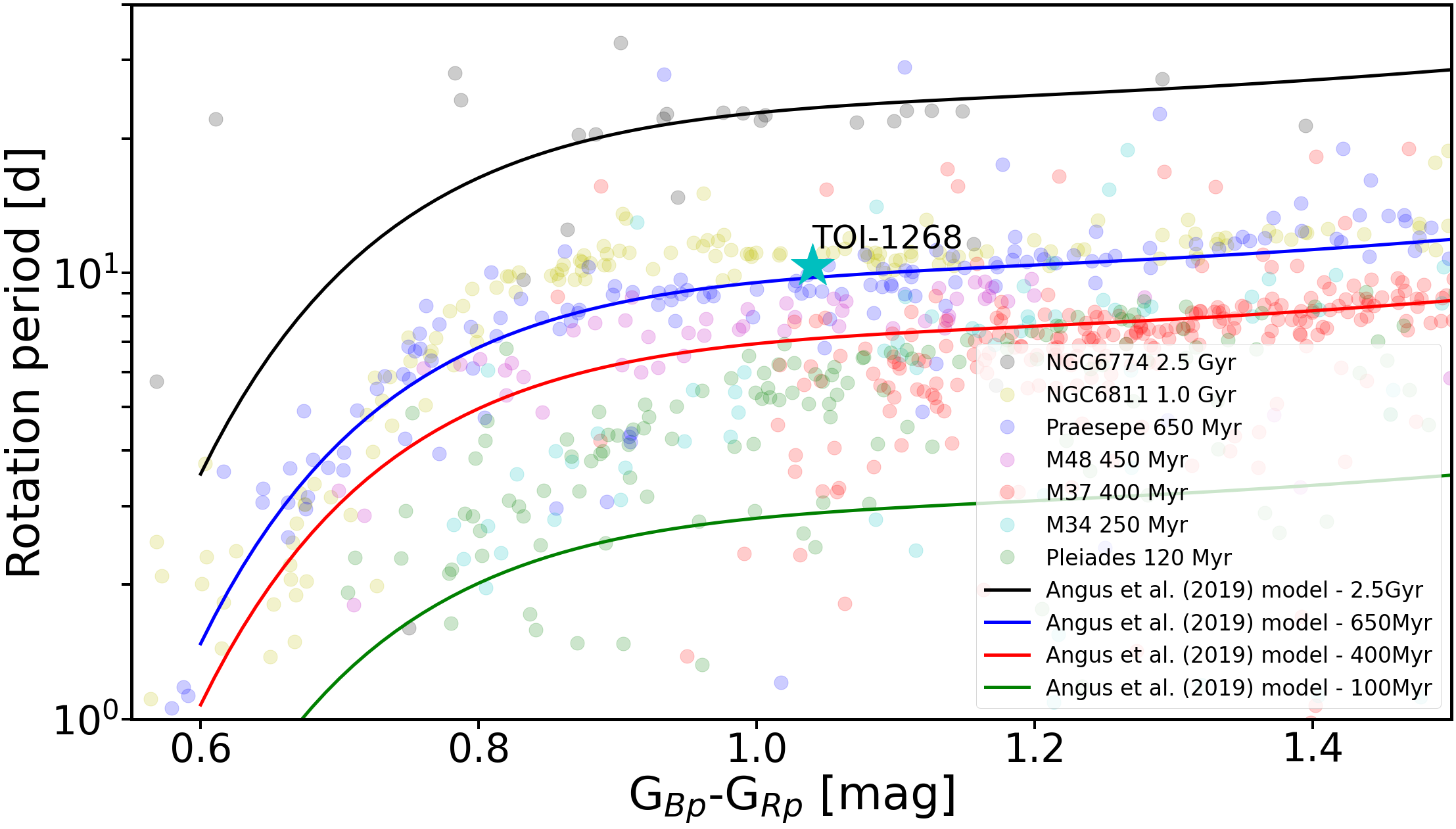}
\caption{Gaia colour vs. rotation period diagram for various members of individual clusters. Lines represent the empirical relation derived in \cite{Angus2019}. TOI-1268 is indicated by a cyan star.} \label{fig:Prot}
\end{figure}

Gyrochronology uses the age-rotation relation to determine the ages of stars, as observations of clusters reveal that rotation slows down as stars become older. Hence, we compare the rotation period vs. colour of TOI-1268 with members of some well-defined clusters: M~35  \citep[$\sim$150 Myr;][]{Meibom09}, M~34  \citep[$\sim$220 Myr;][]{Meibom11}, M~37  \citep[$\sim$400 Myr;][]{Hartman09}, M~48  \citep[$\sim$450 Myr;][]{Barnes15}, Praesepe  \citep[$\sim$650 Myr;][]{Douglas17}, NGC~6811  \citep[$\sim$1 Gyr;][]{Curtis19}, and NGC~6774  \citep[$\sim$2.5 Gyr;][]{Gruner20}. We use the value of TOI-1268's rotation period measured in Sect. \ref{sec:rotation_period}. In Fig. \ref{fig:Prot}, we overplot TOI-1268 on the Gaia colour vs rotation period diagram with cluster members and with curves representing the gyrochronology relation from \cite{Angus2019}. This empirical relation was derived from observations of rotation periods in the Praesepe cluster. According to the rotation period, TOI-1268 has an age between that of Praesepe and NGC~6811, $\sim$650--1000\,Myr.

\subsection{$R'_{HK}$ index}
We used standard relations to convert the time-averaged $S$-index measurements from our spectroscopy into a time-averaged measurement of the activity index, $\log R'_{HK} = -4.41 \pm 0.02$. According to the empirical age-activity relations of \citet{Mamajek:2008}, we then infer an activity age of $330 \pm 50$~Myr. The empirical activity-rotation relations of the same authors then predict a stellar rotation period of $9.7 \pm 1.2$~d for this level of activity, which is consistent with the rotation period of $11.6^{+4.9}_{-2.9}$\,days inferred from the spectroscopic $v\sin i$ together with the stellar radius and also consistent within $1\sigma$ with the rotation period of $10.9 \pm 0.5$~d from the TESS lightcurve. 

\subsection{Lithium Equivalent Width}\label{sec:lithium}

\begin{figure}[!ht]
\centering
\includegraphics[width=0.48\textwidth, trim= {0.0cm 0.0cm 0.0cm 0.0cm}]{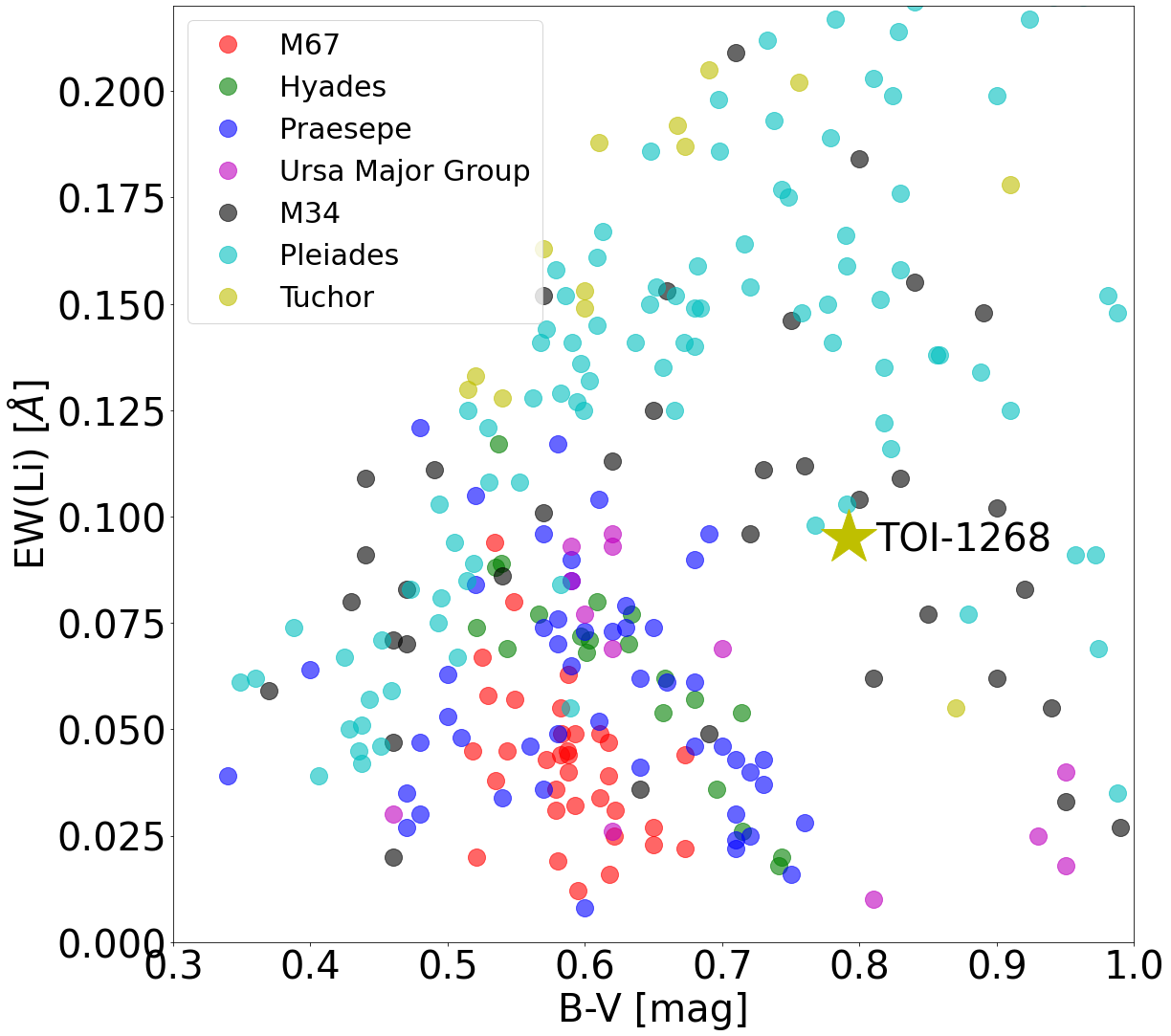}
\caption{Lithium equivalent width vs B--V colour diagram. Points represents members of individual clusters categorized by their colours. TOI-1268 is plotted as a gold star.} \label{fig:Li}
\end{figure}

We used the equivalent width of the lithium line as another age indicator. Lithium is known to be destroyed in the stellar interior through proton capture reactions, and the EW--age relation was confirmed by observations of many clusters. We measured the equivalent width of the lithium line Li\,6708\,$\AA$ with {\tt iSpec}. To do so, we applied a procedure where the line is fitted with a Gaussian profile and the EW corresponds to the area within the gaussian fit. We compare the EW of Li vs $B-V$ colour to members of well-studied clusters in Fig. \ref{fig:Li}. We use data of Tuc--Hor young moving group \citep[$\sim$45 Myr;][]{Mentuch08}, the Pleiades \citep[$\sim$120 Myr;][]{Soderblom93,Lodieu07,Dahm15,Lodieu2019}, M34 \citep[$\sim$220 Myr;][]{Jones97}, Ursa Major Group \citep[$\sim$400 Myr;][]{Soderblom1993}, Praesepe \citep[$\sim$650 Myr;][]{Soderblom993,Lodieu2019}, Hyades \citep[$\sim$650 Myr;][]{Soderblom90,Martin18,Lodieu18,Lodieu19}, and M67 clusters \citep[$\sim$4 Gyr;][]{Jones99}. The data for each cluster taken from the first cited papers are plotted in Fig. \ref{fig:Li} together with the Li EW vs $B-V$ colour of TOI-1268 (see Table \ref{table:system_par}). According to the Li EW, TOI-1268 has an age consistent with the Pleiades and M34. For the age of Pleiades, we adopt the age of 110--150\,Myr from the Li depletion boundary \citep{Barrado14} and for the M34 cluster, we adopt the age of 180--320\,Myr based on \citet{James10} and consistent with \citet{Jones97}. Therefore, based on the Li EW, we assign an age interval of 110--320\,Myr.

\begin{figure*}[!ht]
\centering
\includegraphics[width=1.0\textwidth]{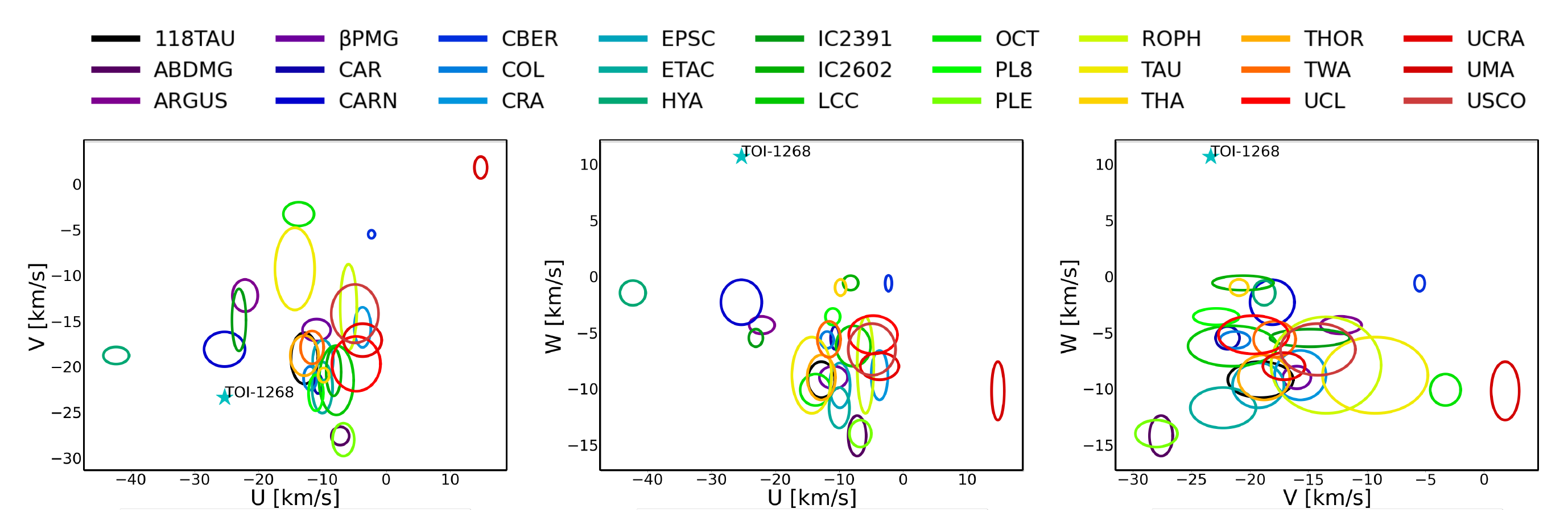}
\caption{U, V, W plots for young stellar associations and TOI-1268. Elipses represent the 1-sigma position of young stellar associations in space velocities U, V, W taken from \cite{Gagne18}. TOI-1268 is plotted as a magenta star.} \label{fig:UVW}
\end{figure*}

\subsection{Membership to young associations}

We used {\tt BANYAN\,$\Sigma$} \citep{Gagne18} to derive TOI-1268's membership probability to young associations within 150\,pc. {\tt BANYAN\,$\Sigma$} is a Bayesian analysis tool that includes 27 young associations with ages in the range between 1--800\,Myr. In addition to {\tt BANYAN\,$\Sigma$}, we also used the kinematic membership analysis code LocAting Constituent mEmbers In Nearby Groups \citep[{\tt LACEwING};][]{Riedel17}. {\tt LACEwING} calculates membership probabilities in 13 young massive groups (YMGs) and three open clusters within 100\,pc. All YMGs are in common with \cite{Gagne18}. As input for both codes, we used astrometric data from $Gaia$ eDR3 listed in Table \ref{table:system_par}. Both codes reveal that TOI-1268 is a field star and is not a member of any young association. In Fig. \ref{fig:UVW}, we plot the 1-sigma position of young stellar associations in space velocities U, V, W taken from \cite{Gagne18} together with TOI-1268. To determine the U, V, W velocities of TOI-1268, we used the python package {\tt PyAstronomy}, specifically the {\tt gal\_uvw}\footnote{\url{https://pyastronomy.readthedocs.io/en/latest/pyaslDoc/aslDoc/gal_uvw.html}} function and derived U\,=\,-25.3\,km/s, V\,=\,-23.4\,km/s, and W\,=\,10.7\,km/s. Visual inspection of the figure also did not confirm the membership to any YMGs; however, the system lies in the region typical of the thin/young disk.

\subsection{Age summary}\label{age_sum}

TOI-1268 looks young (<1 Gyr) according to the majority of age indicators. Stellar isochrones are not very reliable because they provide only very wide estimates for main-sequence stars, and stars of TOI-1268's age have already reached the main sequence, as predicted by stellar models \citep{Cardona21}. Two of the used activity indicators suggest an age of the system below $\sim$400\,Myr. The only age indicator which is not consistent with this age is gyrochronology. The star appears to rotate slower than expected, resulting in a derived age of about 650--1000\,Myr. Looking at the cluster members in Fig. \ref{fig:Prot} we can see that the expected rotation period for a star younger than $\sim$400\,Myr with the corresponding Gaia colour is below six days, hence much below the observed rotation period. We can observe some scatter for each colour and some outliers as the rotation period depends on the initial angular momentum and level of activity; however, TOI-1268's rotation period is significantly higher than is the rotation period observed in clusters of a given age. 

The inconsistency between ages derived from gyrochronology and R'$_{HK}$ index implies that the star either (1) rotates too slowly for its age derived from the R'$_{HK}$ index, or (2) is too active for its age derived from the rotation period. \citet{Mamajek:2008} presents the empirical relation between the R'$_{HK}$ index and P$_{rot}$ through the Rossby number $R_o=P_{rot}/\tau_c$, where $\tau_c$ is the convective turnover of stars. The empirical relation demonstrates that the R'$_{HK}$ index for solar-type dwarfs decay as the Rossby number increases. We can use this empirical relation to predict the rotation period of TOI-1268 from its level of activity. The predicted rotation period is $9.7\pm1.2$\,days, which is consistent with the value derived from photometry. However, it suggests that the observed rotation period is expected for the star's activity level. In such a case, one would expect similar ages predicted from R'$_{HK}$ index and gyrochronology. It is not clear what causes this inconsistency. \citet{Mamajek:2008} derived two different ages from the stellar activity for a large number of solar-type dwarf within 16\,pc. The first one directly uses the empirical relation between age and activity, and the second uses the empirical relation between activity and stellar rotation followed by gyrochronology. These ages are inconsistent for a large number of stars. Hence, in this context, inconsistent ages derived from gyrochronology and R'$_{HK}$ for TOI-1268 look less surprising. The equivalent width of the lithium strongly favour the younger age as lithium of K-dwarf stars is expected to be depleted in Praesepe and Hyades clusters \citep{Soderblom90,Soderblom993,Cummings17}. We adopted the more conservative age of TOI-1268 between 110--1000\,Myr from all age indicators; however, given the majority of age indicators, we can also define a less conservative age of 110--380\,Myr (see Table \ref{table:age}). In upcoming analyses, we either use the broad range or discuss how the results would change considering more and less conservative intervals.

\begin{table}
 \centering
 \caption[]{Summary of age determinations of TOI-1268.
 }
 \label{table:age}
	\begin{tabular}{lcccr} 
		\hline
		\hline
		Technique       & System Age\\
		\hline
		Isochrones & $3.6 \pm 3.5$\,Gyr \\
        Gyrochronology & 650--1000\,Myr (Praesepe--NGC6811) \\
        Lithium EW & 110--320\,Myr (Pleiades--M34) \\
        $R'_{HK}$ & $330 \pm 50$\,Myr \\
        Membership to YMGs & young disk \\
        \hline
        Narrow range & 110--380\,Myr \\
        Broad range & 110--1000\,Myr \\
		\hline
		\hline
	\end{tabular}
\smallskip\\
\end{table}

%
%

\section{Analysis and results}\label{analysis}
\subsection{Frequency Analysis and Stellar Activity}\label{sec:frequency}

\begin{figure}[!ht]
\centering
\includegraphics[width=0.48\textwidth, trim= {0.0cm 0.0cm 0.0cm 0.0cm}]{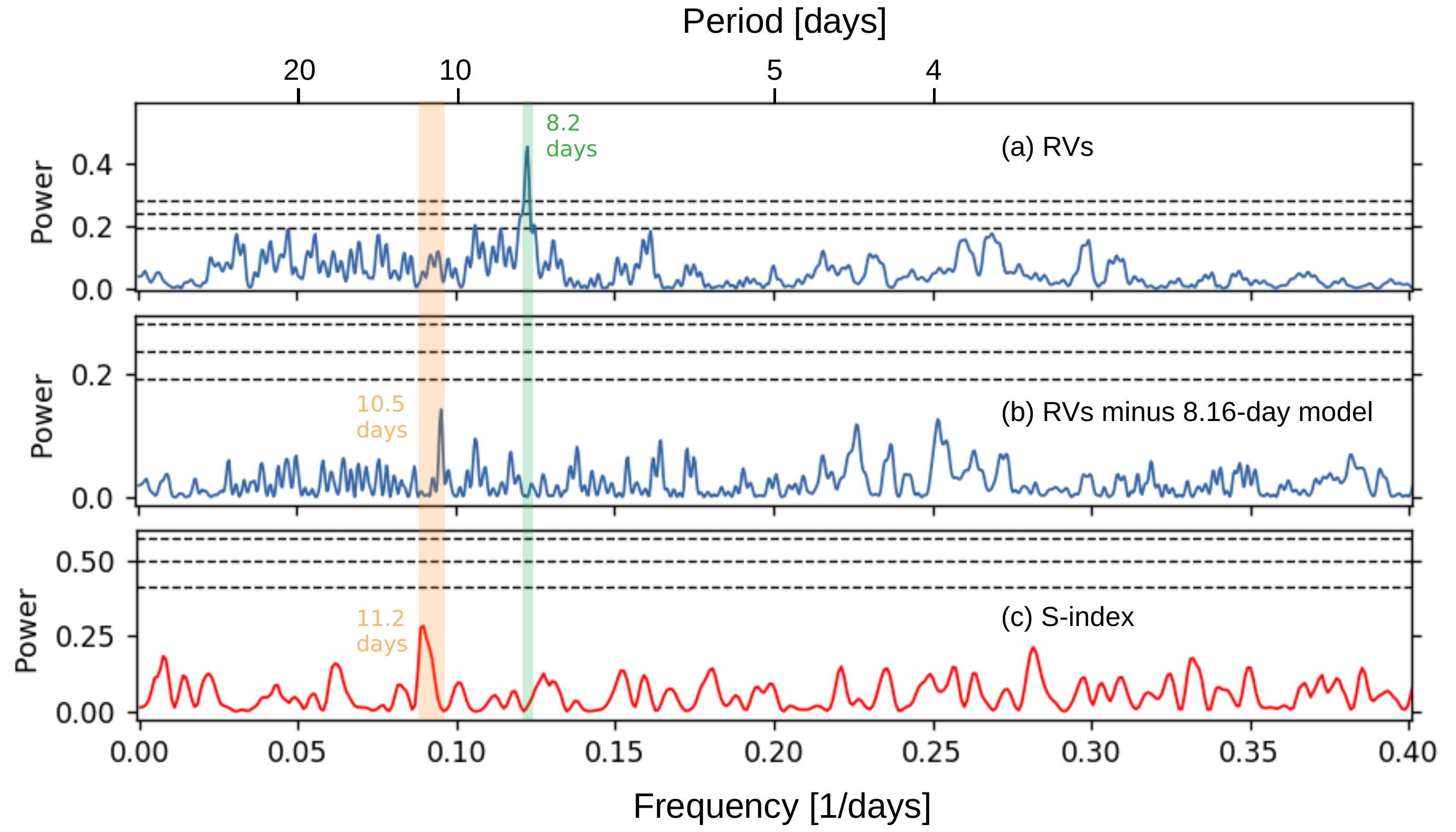}
\caption{Generalised Lomb-Scargle periodograms of RVs (blue) and S-index activity indicators (red) of TOI-1268: (a) Tull RVs, (b) Tull RVs minus 8.16-day model, (c) Tull S-index activity indicator. The vertical green line highlights the orbital period of the planet and the orange region highlights the position of stellar rotation period. Horizontal dashed lines show the theoretical FAP levels of 10\%, 1\%, and 0.1\% for each panel.} \label{fig:freq_analysis}
\end{figure}

In order to distinguish between the Doppler reflex motion induced by the planetary candidates and stellar activity and unveil the presence of possible additional signals, we performed a frequency analysis of the RVs and S-index activity indicator measured from Tull spectra. All spectra used to determine RVs of TOI-1268 were calibrated with iodine cell; hence we can only use activity indicators uncontaminated by iodine lines. We calculated the GLS periodograms \citep{Zechmeister09} of the available time series and computed the theoretical 10 \%, 1 \%, and 0.1 \% false alarm probability (FAP) levels (Fig. \ref{fig:freq_analysis}). The baseline of our observations is about 500 days, corresponding to a frequency resolution of about $1/500=0.002\,d^{-1}$. The most significant period is 8.2\,days, which is consistent with the period of transits from TESS photometry. After fitting for this signal, we do not observe any additional significant peak in the periodogram of residuals. In the periodogram of the S-index, we do not see any significant peak. The only thing worth mentioning is the peak at 11.2 days. The fact that the peak is consistent with the star's rotation period from photometry and is relatively isolated makes it more significant, and we interpret it as the star's rotation period.

%
%

\subsection{Joint RVs and transits modelling}\label{sec:joint_fit}

To simultaneously model transits and RVs we used the {\tt juliet} package. {\tt juliet} is a fitting routine described in \citet{Espinoza19} built on several tools: the transit fitting part is based on \citep[{\tt batman};][]{Kreidberg15}, the radial velocity modelling part uses the package \citep[{\tt radvel};][]{Fulton18}, and GPs modelling uses \citep[{\tt george \& celerite};][]{Ambikasaran15,Foreman17a}. The code performs model comparison via Bayesian evidences and explores the parameter space using nested sampling algorithms included in MultiNest \citep{Feroz09} via the dynesty package \citep{Speagle20}.

\begin{figure*}[!ht]
\centering
\includegraphics[width=1.0\textwidth]{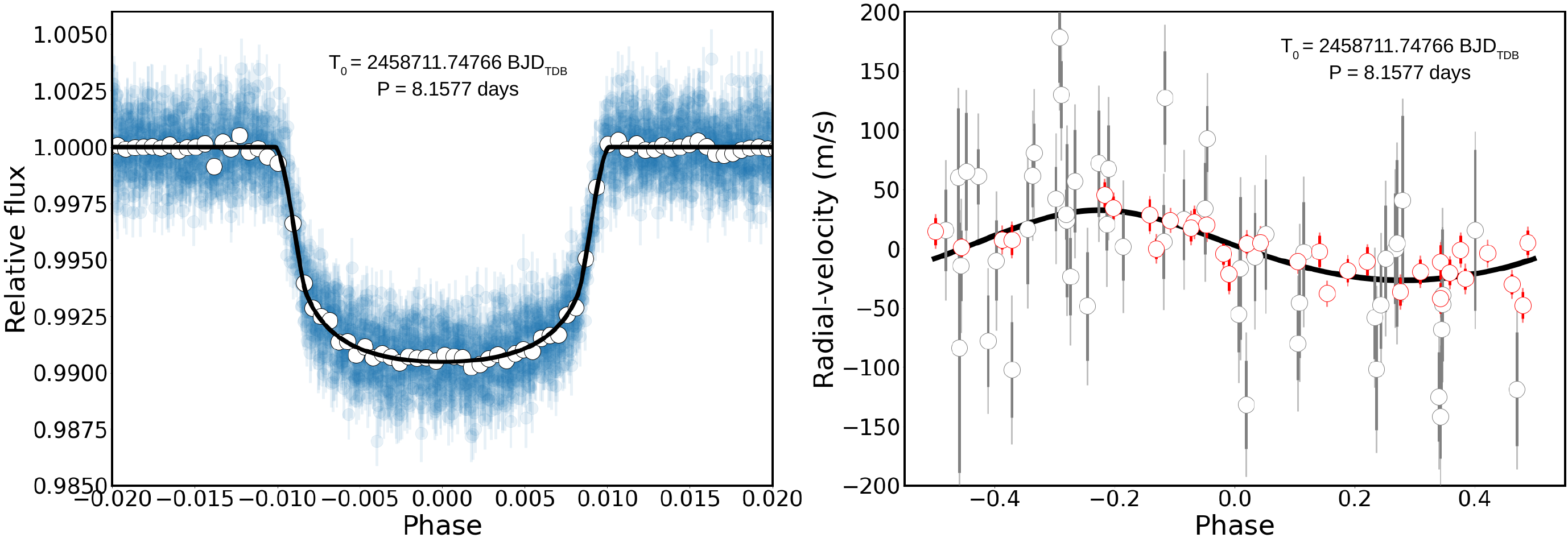}
\caption{Left: The transit light curve of the TOI-1268, fitted with the {\tt juliet} as part of the joint analysis described in Section \ref{sec:joint_fit}. The blue points represent TESS data together with their uncertainties, and the white points are TESS binned data. The black line represents the best transit model. Right: The orbital solution for TOI-1268 showing the {\tt juliet} RV model in black. Grey points represent TCES data and red points Tull data together with their uncertainties and extra jitter term plotted with the lighter grey or red, respectively.} \label{fig:rv_transit}
\end{figure*}

We performed a joint fit using the TESS photometry together with Tull and TCES extracted RVs. We used already corrected LCs for variability with the procedure described in Section \ref{sec:tess_photometry}. We did not include OES RVs in the final analysis as data do not reach sufficient precision to detect TOI-1268b with an average measurement error of 160\,m/s. We used them, however, as the first check to reject an eclipsing binary scenario. {\tt juliet} uses parametrisation from \citet{Espinoza18}. Instead of fitting for the impact parameter $b$ and planet-to-star ratio $p$, it considers parameters $r_1$ and $r_2$ which can be transported to the $b$ and $p$ through the equations found in \citet{Espinoza18}. We also used the quadratic limb darkening law with coefficients $q_1$ and $q_2$. The sampling of the limb darkening coefficients follows the method described in \citet{Kipping13}. Table \ref{tab5} shows priors and posteriors of fitted parameters and Table \ref{table:planet_par} shows transit parameters and physical parameters derived from Table \ref{table:stellar_par}, \ref{tab5} respectively. Phased RVs with RV model together with phased LC and transit model are shown on Fig. \ref{fig:rv_transit}. We detected a Saturn-mass planet with the RV semi-amplitude of $K\,=\,30\pm3\,m/s$ and the transit depth of $\delta\,=\,8186\pm205\,ppm$.

We found a slightly eccentric orbit of e\,=\,$0.092^{+0.035}_{-0.030}$. To investigate the significance of non-circular orbit, we computed the Bayesian model log evidence (ln\,Z) using the {\tt dynesty} package to compare models with fixed zero eccentricity and eccentricity as a free parameter. If the difference in ln\,Z between the models is smaller than two, then they are indistinguishable \citep{Trotta08}. We found $\triangle\,ln\,Z\,=\,0.3$, and hence, we favour a circular orbit scenario. However, we cannot rule out that the orbit is slightly eccentric.

As an independent check of derived parameters, we performed a joint fit with the {\tt pyaneti} package. {\tt pyaneti} is a PYTHON/FORTRAN fitting software described in \citet{Barragan22} that estimates parameters of planetary systems using Markov chain Monte Carlo (MCMC) methods based on Bayesian analysis. We set uniform priors for all fitted parameters following the procedure from \citet{Barragan16}. As an input stellar parameters we used that derived from the HIRES spectrum using the {\tt SpecMatch} software as the stellar mass and radius derived from the Parsec isochrones leads to the mean stellar density of $\rho_\star$\,=\,$2.11 _{ - 0.16 } ^ { + 0.17 }$ ${\rm g\,cm^{-3}}$ which is inconsistent with the value derived from transits as described in \citet{Winn10}. It is not clear why this is the case, and further investigation is needed. The medians and 1-$\sigma$ uncertainties of the fitted and derived parameters together with the stellar input parameters are listed in Table \ref{table:planet_par_pyan}. The correlations between the free parameters from the MCMC analysis and the derived posterior probability distributions are shown in Fig. \ref{fig:correlation}. All fitted parameters are in good agreement with those derived with {\tt juliet}.

\begin{table*}
	\centering
	\caption{Fitted parameters from {\tt juliet} analysis.}
	\label{tab5}
	\begin{tabular}{lccccccr}
		\hline
		\hline
		Parameter       & Units & Value & Priors\\
		\hline
		\hline
		Stellar parameters: & & \\
        $\rho_\star$\dotfill & Density (kg~m$^{-3}$)\dotfill & $1358^{+258}_{-298}$ & $\Im$(100, 10000) \\
        \smallskip\\
        Planet parameters: & & \\
        $P$\dotfill & Period (days)\dotfill & $8.1577080^{+0.0000044}_{-0.0000041}$ &  $\mathcal{N}$(8.15,0.10)\\
        $T_0$\dotfill & Optimal conjunction Time (\bjdtdb)\dotfill & $2458711.74766^{+0.00020}_{-0.00024}$ & $\mathcal{N}$(2458711.75,0.10)\\
        $r1$\dotfill & Parametrisation for $p$ and $b$\dotfill & $0.537^{+0.057}_{-0.072}$ & $\mathcal{U}$(0, 1)\\
        $r2$\dotfill & Parametrisation for $p$ and $b$\dotfill & $0.08914^{+0.00120}_{-0.00094}$ & $\mathcal{U}$(0, 1)\\
        $e$\dotfill & Eccentricity\dotfill & $0.09^{+0.04}_{-0.03}$ & $\mathcal{U}$(0, 0.5) \\
        $\omega$\dotfill & Argument of periastron\dotfill & $-22^{+57}_{-34}$ & $\mathcal{U}$(-180, 180)\\
        $K$\dotfill & RV semi-amplitude (m/s)\dotfill & $30\pm3$ &  $\mathcal{U}$(0, 100)\\
        \smallskip\\
        Photometry parameters: & & \\
        $\sigma_{TESS}$\dotfill & Extra jitter term (ppm)\dotfill & $2^{+8}_{-2}$ &  $\Im$(0.1, 1000)\\
        $D_{TESS}$\dotfill & Dilution factor\dotfill & $1.0$ &  fixed\\
        $M_{TESS}$\dotfill & Relative flux offset\dotfill & $0.0000008^{+0.0000039}_{-0.0000041}$ & $\mathcal{N}$(0,0.1)\\
        $q_{1,TESS}$\dotfill & Quadratic limb darkening parametrisation\dotfill & $0.70^{+0.20}_{-0.17}$ & $\mathcal{U}$(0, 1)\\
        $q_{2,TESS}$\dotfill & Quadratic limb darkening parametrisation\dotfill & $0.14^{+0.06}_{-0.05}$ & $\mathcal{U}$(0, 1)\\
	    \smallskip\\
        RV parameters: & & \\
        $\sigma_{Tull}$\dotfill & Extra jitter term (ms$^{-1}$)\dotfill & $9\pm3$ &  $\Im$(0.01, 100)\\
        $\sigma_{TCES}$\dotfill & Extra jitter term (ms$^{-1}$)\dotfill & $48\pm7$ &  $\Im$(0.01, 100)\\
        $\mu_{Tull}$\dotfill & Systemic velocity (ms$^{-1}$)\dotfill & $-47224\pm2$ &  $\mathcal{U}$$(-47200, -47250)$\\
        $\mu_{TCES}$\dotfill & Systemic velocity (ms$^{-1}$)\dotfill & $-139\pm7$ &  $\mathcal{U}$$(-150, -100)$\\
        \smallskip\\
		\hline
		\hline
	\end{tabular}
\smallskip\\
\end{table*}

\begin{table}
 \centering
 \caption[]{Derived parameters from {\tt juliet} analysis.
 }
 \label{table:planet_par}
	\begin{tabular}{lcccr} 
		\hline
		\hline
		Parameter       & TOI-1268b\\
		\hline
		\multicolumn{3}{c}{Derived parameters} \\
        $R_p/R_\star$\dotfill & Radius of planet in stellar radii\dotfill & $0.08914^{+0.00120}_{-0.00094}$ \\
        $a/R_\star$\dotfill & Semi-major axis in stellar radii\dotfill & $16.84^{+1.01}_{-1.33}$ \\
        $a$\dotfill & Semi-major axis (AU)\dotfill & $0.072^{+0.009}_{-0.010}$ \\
        $b$\dotfill & Transit Impact parameter\dotfill & $0.306^{+0.085}_{-0.108}$ \\
        $i_p$\dotfill & Inclination ($^\circ$)\dotfill & $88.98^{+0.38}_{-0.34}$ \\
        $\delta$\dotfill & Transit depth (fraction)\dotfill & $0.0080^{+0.0002}_{-0.0002}$ \\
        \smallskip\\
        \multicolumn{3}{c}{Derived physical parameters} \\
        $M_p$\dotfill & Mass ($M_\oplus$)\dotfill & $102\pm11$ \\
        $M_p$\dotfill & Mass ($M_J$)\dotfill & $0.29\pm0.04$ \\
        $R_p$\dotfill & Radius ($R_\oplus$)\dotfill & $9.0\pm0.7$ \\
        $R_p$\dotfill & Radius ($R_J$)\dotfill & $0.82\pm0.06$ \\
		\hline
		\hline
	\end{tabular}
\smallskip\\
\end{table}

\section{Discussion}\label{sec:discussion}

\subsection{Mass-Radius diagram}

\begin{figure*}[!ht]
\centering
\includegraphics[width=1.0\textwidth]{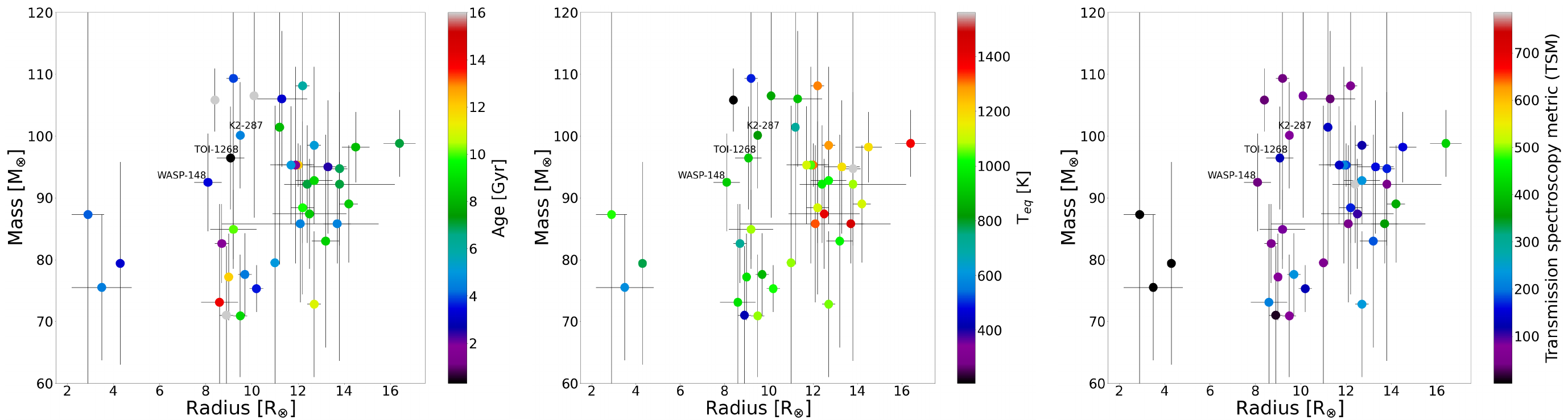}
\caption{The population of known planets within the mass interval 70--110\,$M_{\oplus}$. The position of TOI-1268b is highlighted together with two planets sharing similar properties. The left subplot is coloured with respect to the age of the systems (the grey colour is used for systems without an age estimate in literature), the middle one with respect to the equilibrium temperature of the planets, and the right one with respect to the transmission spectroscopic metric.} \label{fig:population}
\end{figure*}

By combining data from the TESS mission and ground-based spectroscopy, we have confirmed the planetary nature of a $P\,=\,8.16$\,days candidate around the V\,=\,10.9 mag K-type star TOI-1268. We found that the physical parameters of TOI-1268b ($M_p$\,=\,$0.303\pm 0.026$\,$M_J$, $R_p$\,=\,$0.81\pm 0.05$\,$R_J$) are consistent with those of Saturn. We compare the mass and radius of TOI-1268b with the population of known planets within the mass interval 70--110\,$M_{\oplus}$ in Fig. \ref{fig:population}. Three subplots are created, in which each colour represents a different parameter: age, equilibrium temperature, and transmission spectroscopic metrics. 

The most prominent group in Fig. \ref{fig:population} is the group of inflated planets with a radius of about 12 $R_{\oplus}$ or larger and equilibrium temperatures higher than 1000\,K. TOI-1268b is in the second group of non-inflated planets with two members that share similar bulk properties: WASP-148b \citep{Hebrard20} and K2-287b \citep{Jordan19}. The non-inflated structure of TOI-1268b is expected given its relatively low equilibrium temperature of $T_{eq}$\,=\,919\,K at which the inflation mechanism of hot Jupiters does not play a significant role \citep{Kovacs10,Demory11}. The equilibrium temperature is computed according to the equation in \citet{Kempton18} considering zero albedo and full day-night heat redistribution. Both WASP-148b and K2-287b also have equilibrium temperatures below 1000\,K, a rough theoretical lower limit for planet inflation. TOI-1268b appears to be a very interesting target to explore the drivers of atmospheric inflation as it seems to be just below the threshold.

\citet{Kempton18} proposed a metric that can be used to evaluate the suitability of planets for further atmospheric characterisation via transmission spectroscopy study. The right subplot in Fig. \ref{fig:population} is coloured according to this metric. We can see that TOI-1268b is one of the best candidates for the atmospheric characterisation among non-inflated Saturn-mass planets. 

\subsection{Tidal interaction}

Giant planets, such as TOI-1268b, are thought to have formed at larger separations from their host stars and to have migrated inwards through interaction with the proto-planetary disc \citep{Coleman:2017}. At close distances from the star, tidal interactions play a significant role and circularise the planetary orbit on timescales that can be calculated. There are several reasons to study tidal interactions. First, the circularisation timescales depend on the tidal quality factors of the planet and the star, which are difficult to measure. As we will discuss in this section, systems such as TOI-1268 can put some constraints on these parameters. Second, if we observe an eccentric planet in an old system, where one would have expected the orbit of the planet to be already circularised, we can suspect additional interactions from another planet or a distant companion. 

\begin{figure}[!ht]
\centering
\includegraphics[width=0.48\textwidth, trim= {0.0cm 0.0cm 0.0cm 0.0cm}]{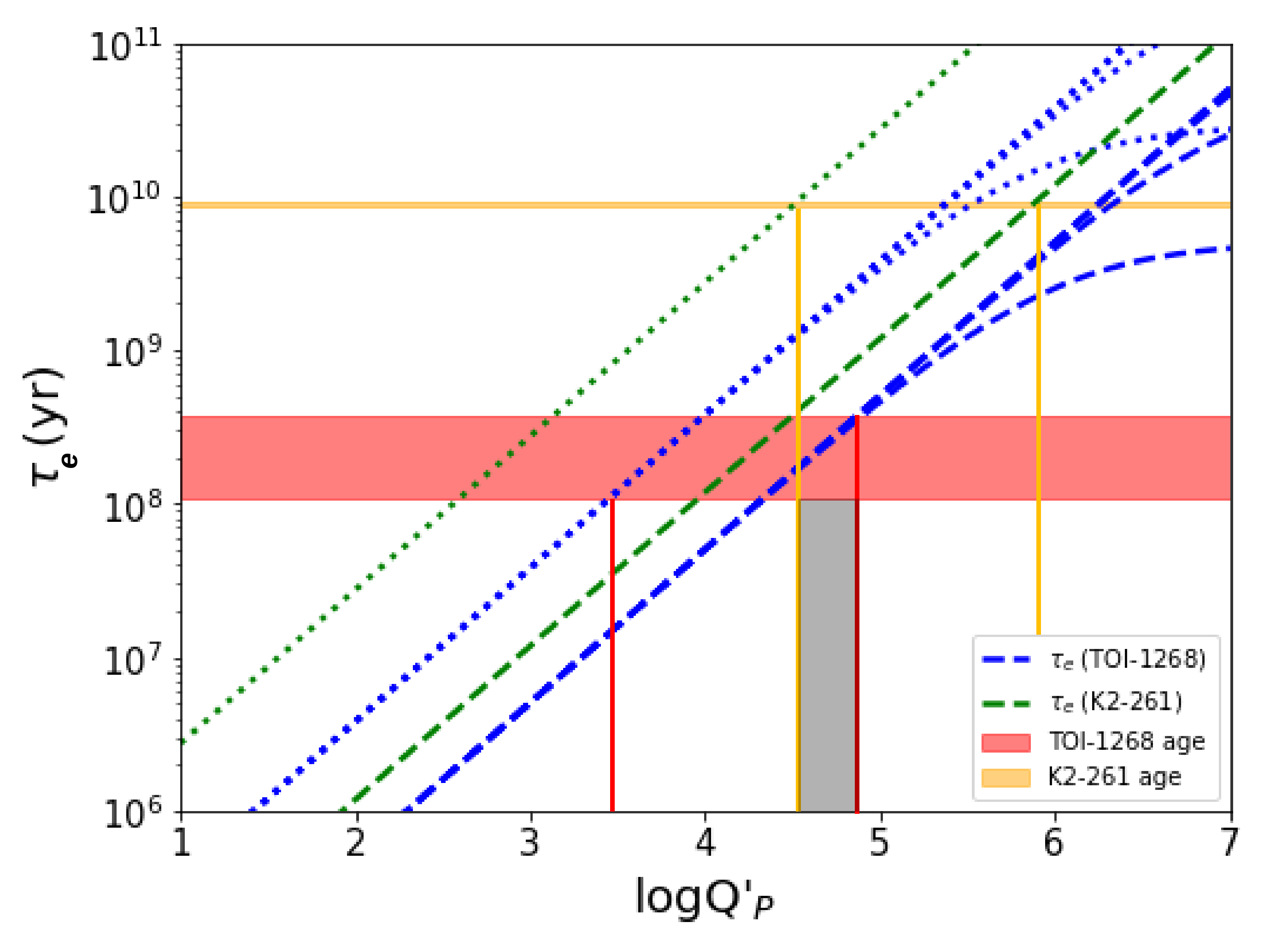}
\caption{Values of the tidal circularisation timescales as a function of planet tidal quality factor for TOI-1268b (blue lines) and K2-261 (green lines). Dashed and dotted lines of each colour represent the lower and upper boundaries of plausible values of timescales based on uncertainties of parameters used in Equation \ref{eqn:tidecirc}. Curved blue lines represent the dependence on stellar tidal quality factors for $Q'=10^{4}, 10^{5}, 10^{6}, 10^{7}, 10^{8}$, respectively. It demonstrates that the stellar tidal quality factor influences the circularisation timescale only if the value of $Q'_{\mathrm{p}}$ is larger than $Q'=10^{5.5}$. The horizontal lines show the systems' ages. Vertical lines then highlight the values of $Q'_{\mathrm{p}}$ in positions where timescales cross the ages of the systems. These positions define the lower limit for $Q'_{\mathrm{p}}$ in the case of K2-261 and the upper limit in the case of TOI-1268. The grey region then represents an overlap in $Q'_{\mathrm{p}}$ between the lower limit and upper limit.} \label{fig:tidal}
\end{figure}

According to \cite{Jackson2008}, the timescale for orbital circularisation  for a close-in companion is:

\begin{equation}
\label{eqn:tidecirc}
    \frac{1}{\tau_e}=\bigg[\frac{63}{4}\sqrt{GM_{\star}^3}\frac{R_{\mathrm{p}}^5}{Q'_{\mathrm{p}}M_{\mathrm{p}}}+\frac{171}{16}\sqrt{G/M_{\star}}\frac{R_{\star}^5M_{\mathrm{p}}}{Q'_{\star}}\bigg]a^{-\frac{13}{2}},
\end{equation}

where $Q'_{\star}$ and $Q'_{\mathrm{p}}$ are the tidal quality factors of the star and planet, respectively. The tidal quality factor $Q$ is a parametrisation of the response of the body's interior to tidal perturbation, which is defined as:
\begin{equation}
\label{eqn:tideQ}
    Q^{-1}\equiv\frac{1}{2{\pi}E_0}\oint\bigg(-\frac{dE}{dt}\bigg)\,dt,
\end{equation}

where $E_{0}$ is the peak energy stored in the tidal distortion during the cycle, dE/dt is the rate of dissipation and its integral defines the energy lost during the cycle \citep{Goldreich66}. The tidal quality factor then represents the ratio of the elastic energy stored to the energy dissipated per cycle of the tide 
--  larger values of tidal factors lead to longer circularisation timescales. The tidal quality factor in Equation \ref{eqn:tidecirc} further includes the Love number, the correction factor for the tidal-effective rigidity of the body and its radial density distribution, $Q'=3Q/2k_{2}$ \citep{Goldreich66,Jackson2008}. For an homogeneous fluid body $k_{2}=3/2$. The value of $Q'$ is difficult to estimate, and possible values span large intervals from $10^2$ for rocky planets \citep{Clausen15}, to $10^{5-6}$ for some giant planets \citep{Lainey09}, up to $10^{8-9}$ in case of some stars \citep{Collier18}, with typical scatter of several orders of magnitude. To assess the effect of the uncertain values of $Q'_{\star}$ and $Q'_{\mathrm{p}}$ on the circularisation timescale, in Fig. \ref{fig:tidal} we plot this timescale as a function of the tidal quality factors. We also plot the age of TOI-1268, which can be used to constrain some limits. 

Before any conclusions, we need to note that the equation above has two components. The first depends on the tidal quality factor of the planet, and the second on the tidal quality factor of the star. Hence, the values of these factors, together with the planet's mass, will define which component is significant. For rocky planets with low mass and value of $Q'_{\mathrm{p}}\sim10^2$, the first component will always be dominant. For giant planets such as TOI-1268b, the first component is usually dominant, and as we see in Fig. \ref{fig:tidal} the tidal quality factor of the star starts to play a role only for $Q'_{\mathrm{p}}>10^{5.5}$. For brown dwarfs, the second component is usually dominant, and the tidal quality factor of the planet starts to play a role only for $Q'_{\star}>10^6$, as shown in \citet{Subjak20}. It means that while for planets we are usually interested only in $Q'_{\mathrm{p}}$ in order to study tidal interactions, for brown dwarfs, we need to consider both $Q'_{\star}$ and $Q'_{\mathrm{p}}$, complicating the whole process. For Saturn-mass planets, it is sufficient to consider only the first component. Another important thing to note is that circularisation timescale in this form represents the exponential damping of eccentricity on this timescale; hence the time needed to fully circularise orbit is two or three times larger depending on the initial value of the eccentricity. Finally, to justify that we compare circularisation timescales with ages, we assume that the time the planet has spent at a small orbital distance is comparable with the age of the star. In other words, we assume that planets migrate inwards very early in the lifetimes of their systems \citep{Trilling98,Murray98,Suarez21}.

First, we searched for systems hosting non-inflated Saturn-mass planets on eccentric orbits to put constraints on the lower limit of $Q'_{\mathrm{p}}$. The reason why we target only non-inflated planets is that the differences in the internal structure of planets can lead to different values of $Q'_{\mathrm{p}}$. Hence, we assume that non-inflated Saturn-mass planets share a similar value of $Q'_{\mathrm{p}}$. However, such an assumption is not physically justified. Examples of such systems are NGTS-11 \citep{Gill20} or K2-287 \citep{Jordan19}; however, the system with ideal parameters to constrain $Q'_{\mathrm{p}}$ is K2-261 \citep{Johnson18}. K2-261 hosts a planet with mass $M_p$\,=\,$71\pm10\,M_{\oplus}$, radius $R_p$\,=\,$9.5\pm0.3\,R_{\oplus}$, orbital period $P\,=\,11.63344\pm0.00012$\,days and eccentricity $e\,=\,0.39\pm0.15$. Such a high eccentricity means that the age of the system is lower than the circularisation timescale (one exponential damping of eccentricity) calculated for this system, putting the lower limit on the
$Q'_{\mathrm{p}}$. Furthermore, the age of the system, $8.8\pm0.4$\,Gyr, is derived with great precision because the star is located near the main sequence turn-off. The derived lower limit of $Q'_{\mathrm{p}}\,\sim\,10^{4.5}$ can be seen in Fig. \ref{fig:tidal}, where we compare the circularisation timescale of the system with its age. The circularisation timescale is increasing with the value of the tidal quality factor, and the position where it crosses the system's age defines the lower limit for $Q'_{\mathrm{p}}$.

Analogously we can discuss the upper limit of $Q'_{\mathrm{p}}$ looking at systems with planets on circular orbits. Here we can use the TOI-1268 system. Even though we showed in Sec \ref{sec:joint_fit} that the favoured eccentricity is zero, we cannot rule out that the orbit is slightly eccentric. Hence, for this system, we can only expect that the age is larger than the one circularisation timescale -- one exponential damping of eccentricity (for a circular orbit, we can set the limit two or three circularisation timescales). In other words, we expect initial eccentricity after the migration to be equal to or larger than 0.15--0.35. This is supported by the observations of eccentricities in other systems, such as K2-261 or K2-287 with observed eccentricities e\,=\,$0.39\pm0.15$, e\,=\,$0.48\pm0.03$, respectively. Furthermore, according to the Kepler survey, the systems with single transiting planets have a mean eccentricity $\sim0.3$ \citep{Xie16}. TOI-1268 then provide the upper limit on the value of $Q'_{\mathrm{p}}$ which is $Q'_{\mathrm{p}}\,\sim\,10^{4.9}$ as can be seen in Fig. \ref{fig:tidal}. However, if we consider the broad interval for the age (see Section \ref{age_sum}) the upper limit for $Q'_{\mathrm{p}}$ would be $10^{5.3}$. Hence, using lower and upper limits, we found the overlap, which is between $Q'_{\mathrm{p}}\,\sim\,10^{4.5-4.9}$ or $Q'_{\mathrm{p}}\,\sim\,10^{4.5-5.3}$ based on the age of TOI-1268. Various unknowns can affect our results, such as a gravitational interaction with other (undetected) bodies or differences in migration processes and values of initial eccentricities. Even if individual systems can be affected by these unknowns with a large sample of well-characterised Saturn-mass planets, we can obtain better insight into the general value of $Q'_{\mathrm{p}}$ and test the theoretical predictions.

We can now compare these results with the tidal quality factor measured for Saturn in our solar system. In \citet{Lainey12} the $k_2/Q_S = 2.3\pm0.7 \times 10^{-4}$ has been evaluated by studies based on the orbital migration of moons using astrometric data spanning more than a century. Such values are higher than our derived interval of $k_2/Q = 0.2-0.5 \times 10^{-4}$ ($0.1-0.5 \times 10^{-4}$ for older age of TOI-1268). The recent measurements of Titan's orbital expansion rate obtained with the Cassini spacecraft have indicated that Saturn's Q value can be two orders of magnitude lower ($k_2/Q_S$ is then higher) than the value from \citet{Lainey12}. The study by \citet{Lainey20} indicates that resonant tides are important for Saturn, and the resonance locking mechanism can explain the predicted $Q_S$ based on the orbital migration of moons. Hence, comparisons with values measured for Saturn are of limited use.

\subsection{TOI-1268b in the context of young planets}

To date, only six systems with transiting gas giant planets with measured mass are known with age below 1\,Gyr: Kelt-9 \citep{Gaudi17}, Kelt-17 \citep{Zhou16}, WASP-178 \citep{Rodriguez20}, Mascara-4 \citep{Dorval20}, AU Mic \citep{Plavcan20,Martioli21} and V1298\,Tau \citep{Suarez21,Poppenhaeger21}. There are several more systems with measured ages but without masses or only with the upper limits on their masses \citep{Lund17,Rizzuto20}. V1298\,Tau has a young age of 10--30\,Myr and AU Mic has an age of $22\pm3$\,Myr, while Kelt-17, Mascara-4 have larger ages (0.5--0.9\,Gyr). WASP-178 has a wide age estimate of 200--750 Myr, and Kelt-9 has an age of $\sim$300\,Myr. However, Kelt-9, together with three other systems with ages of several hundreds of years, orbits an A-type star, thereby making TOI-1268 quite distinctive and more comparable with the youngest known K dwarf hosting giant planets, V1298\,Tau \citep{David19}.

\begin{figure}[!ht]
\centering
\includegraphics[width=0.48\textwidth, trim= {0.0cm 0.0cm 0.0cm 0.0cm}]{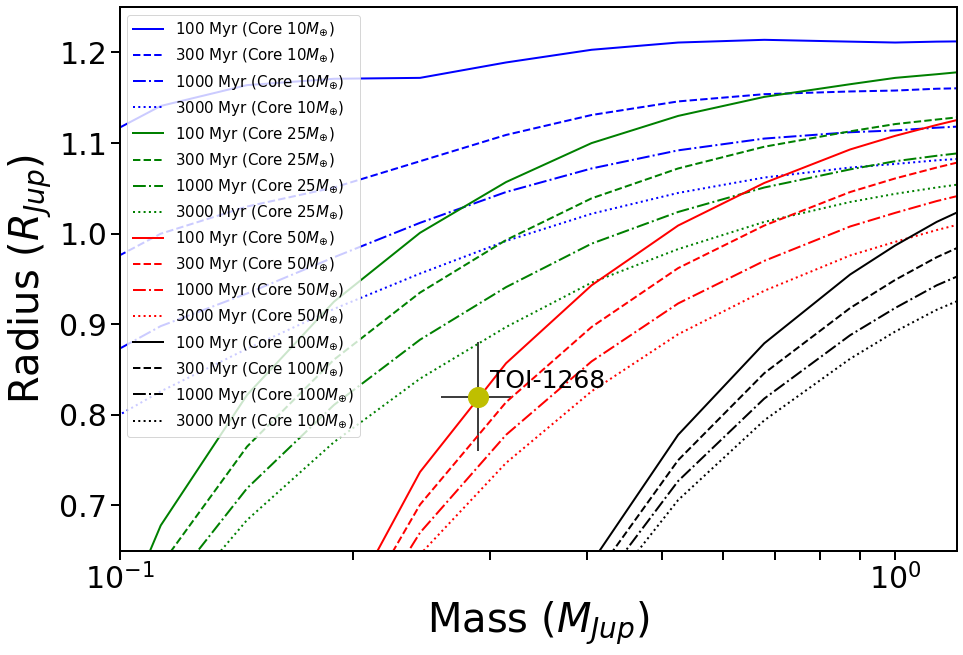}
\caption{The isochrones computed from the thermal evolution model described in \citet{Fortney07} as they depend on the age of the system and the core's mass. The isochrones are adjusted to the TOI-1268's luminosity and TOI-1268b's distance.} \label{fig:evol_model}
\end{figure}

The mass and radius of TOI-1268b are surprisingly similar to that of Saturn; hence density is similar too: $0.71 _{ - 0.13 } ^ { + 0.17 }$\,$g\,cm^{-3}$ vs 0.69 for our Saturn, even though the system is much younger than our Solar system. We used the thermal evolution model described in \citet{Fortney07} to determine the heavy element mass of TOI-1268b. It can be seen in Fig. \ref{fig:evol_model} where we plot the isochrones in the mass-radius plane as they depend on the age of the system and the heavy element mass. The isochrones are adjusted to the TOI-1268's luminosity and TOI-1268b's distance. We found that the heavy elements would need to have a mass of about $50M_{\oplus}$ to explain the observed mass and radius. If we consider for TOI-1268b the heavy element mass around $25M_{\oplus}$ comparable with that of Saturn, the object of TOI-1268b's age should have a larger radius than is measured (see Fig. \ref{fig:evol_model}). It is consistent with the results of V1298\,Tau b and e \citep{Suarez21} that have a mass and a radius comparable with that of Jupiter, even though the system has an age of only 10--30 Myr. \citet{Suarez21} suggest that current models are not able to describe well the contraction of hot giant planets, and the contraction timescale can be much shorter than expected. TOI-1268 is consistent with such a scenario. However, another process that needs to be considered is atmospheric evaporation, which is discussed in the next subsection.

\subsection{Atmospheric erosion}

As we have shown, TOI-1268\,b is a Saturn-mass planet orbiting a young K dwarf, making it a potentially interesting target for the studies of atmospheric erosion. Unfortunately, the flux of the star in the X-ray and extreme UV region is unknown, so it is not yet possible to calculate the expected mass-loss rate of the planet. However, we can estimate them. Using the relation 
$$ L_x\sim (3 \pm 1) \times 10^{28}~t^{-1.5\pm0.3}~
{\rm [erg\,s^{-1}],}$$
with $t$ in Gyrs, we estimate $\log L_x=28.3-30.3$ \citep{Guedel2004}. On the other hand, using the values obtained by \citet{Jackson2012} for stars in clusters of different ages, we find $\log L_x=28.5-30.1$. It is thus reasonable to assume that the $\log L_x \approx 28.5-30.0$.

\begin{figure*}[!ht]
\centering
\includegraphics[width=0.98\textwidth, trim= {0.0cm 0.0cm 0.0cm 0.0cm}]{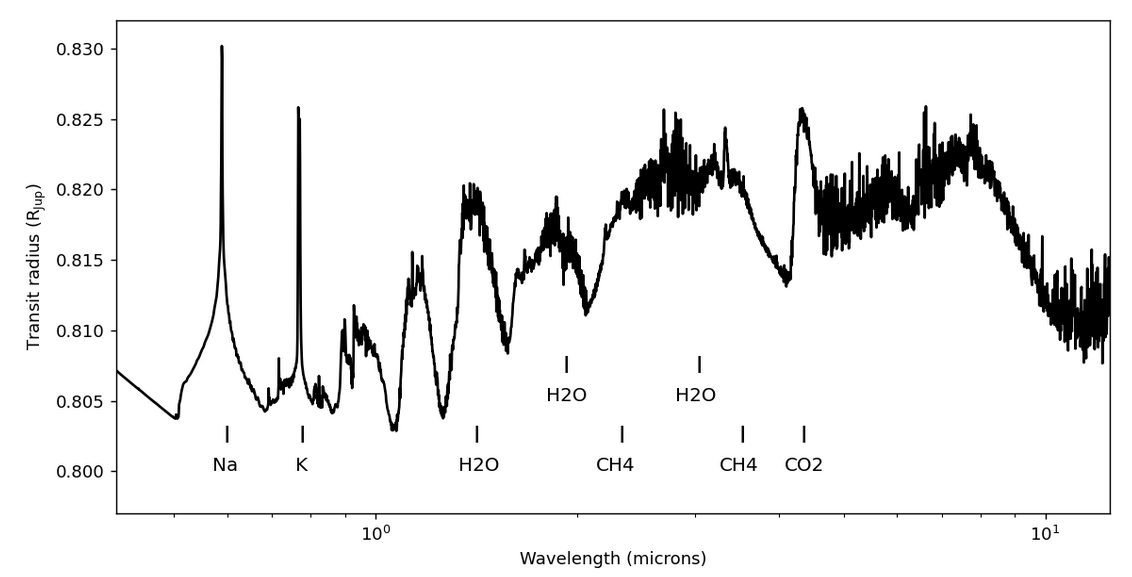}
\caption{Transit radius of the planet as a function of wavelength. The model of the transmission spectrum of TOI-1268b was generated by PetitRADTRANS \citep{petit}. The strongest atmospheric absorbers are marked.} \label{fig:TS}
\end{figure*}

Together with the mass and radius obtained for the planet, this gives a mass-loss range between $10^{11}$ to $10^{12}\, {\rm g\,s^{-1}}$ \citep[using the same relation as in][]{Foster2021}. We used several assumptions mentioned in \citet{Foster2021} and discussed in \citet{Poppenhaeger21}. This mass-loss is rather high \citep{Foster2021}. At this mass-loss rate, the planet would lose about one per cent or less of its mass in 100\,Myrs. Because the mass-loss rate will decrease with time, it is unlikely that TOI-1268\,b will become a rocky planet. However, studying a young planet with a high mass-loss rate is important to understanding the mass-loss of planets in general. This is illustrated by the cases of  V1298\,Tau and GJ\,143.

\citet{Poppenhaeger21} studied the evaporation of the planets of the young K0--1.5 star V1298\,Tau in detail and concluded that the innermost two planets can lose significant parts of their gaseous envelopes and could be evaporated down to their rocky cores, depending on the stellar spin evolution of the star. Complete erosion of the atmospheres of these planets is possible because V1298\,Tau is a young, active star, and the planets are small.

GJ\,143 is a nearby K dwarf \citep{Dragomir2019} of an age $3.8\pm3.7$\,Gyr characterised by $\log\,L_x=27.2$  \citep{Foster2021}. The outer planet GJ\,143\,b has a period of 35.6 days, a mass of $\rm M_\mathrm{p}\,=\,22.7^{+2.2}_{-1.9}\,M_{\oplus}$ \citep{Dragomir2019}, and receives $\rm F_\mathrm{X,pl}\,=\,16\,erg\,s^{-1} cm^{-2}$. Its mass-loss rate  is only $\rm 5\times10^7\,g\,s^{-1}$ \citep{Foster2021}. The inner planet, GJ\,143\,c, has an orbital period of 7.8 days and its mass is below  $\rm 3.7\,M_{\oplus}$ and it receives $\rm F_\mathrm{X,pl}\,=\,123\,erg\,s^{-1} cm^{-2}$. What makes this system interesting is that the inner planet is rocky, and the outer still has a gaseous envelope. It is thus possible that the inner one once also had a gaseous envelope that was eroded when the star was younger. The outer one kept its atmosphere because the erosion was not high enough to fully erode it. The very interesting aspect of this system is that the erosion must have happened when the star had an $\rm L_x$-value which was in the range that we expect for TOI-1268\@. Studying TOI-1268 thus helps to understand the evolution of GJ\,143\@. 

The examples of V1298\,Tau and GJ\,143 show how important the mass loss for the evolution of planets can be. Although we do not witness the transition from a gas giant to a rocky planet in TOI-1268, this object is an interesting case to study the mass loss processes in young planets. TOI-1268 is a transiting planet of a nearby, young star, its mass-loss rate is likely to be high, and we have determined the mass and radius of the planet. These properties make it a good target for future studies of the mass loss of planets. 

Atmospheric mass loss can be constrained by measuring hydrogen and He I lines. The first detection of the evaporating atmosphere has been performed by \citet{vidal} by studying the Lyman-alpha line; however, such observations are hindered by interstellar absorption and geocoronal emission. H-alpha has been detected in atmospheres of hot Jupiters around active K-dwarfs \citep{jensen,chen}. The metastable triplet line can be used to derive mass loss, thermospheric temperature, and atmospheric dynamics through high-resolution transmission spectroscopy \citep{paragas}.

TOI-1268b receives large amount of EUV radiation. We have used relation published in \citet{sree} to obtain $\rm{log(F_{EUV}/F_{bol})}=-4.48$. The derived $F_{EUV}\approx 4500\, \rm{erg\,s^{-1}\,cm^{-2}}$suggests favourable conditions for detection of the He I and hydrogen features as the $F_{EUV}$ is approximately 3 times higher compared to WASP-69b, a planet with similar planetary parameters and strong detected helium features using the CARMENES instrument \citep{nort}. This makes TOI-1268b an excellent candidate to study mass-loss in the future.

As mentioned above, due to its high atmospheric metric TOI-1268b constitute an excellent target for further atmospheric characterisation. The equilibrium temperature of TOI-1268b puts this planet into a class of models for equilibrium temperature of 1000\,K by \citet{Fortney10}. The most prominent features in such an atmosphere are $\rm{H_20}$, Na and K absorption. We created (Fig. \ref{fig:TS}) a simple atmospheric model in {\tt PetitRADTRANS} \citep{petit} of expected transmission spectrum. Such atmospheric features can be detected with current instruments and make TOI-1268b a viable target for future atmospheric studies.
 
\section{Summary}

We have presented an analysis of the non-inflated Saturn-mass planet, TOI-1268b, transiting an early K dwarf star. TOI-1268b has a moderate orbital period ($ P = 8.1577080\pm 0.0000044$ days). The age of the system is estimated to be less than one Gyr using various age indicators, making the planet the youngest Saturn-mass planet known. Given a relatively bright primary, TOI-1268b is a great target for future transmission spectroscopy studies and one of the best from the population of non-inflated Saturn-mass planets. We use system parameters to discuss tidal interactions and constrain the values of a tidal quality factor for non-inflated Saturn-mass planets.\\

TOI-1268b has also been independently confirmed by \citet{Dong22}. The two papers have been prepared and submitted simultaneously and are the result of independent observations and analyses. We are very grateful to Jiayin Dong and collaborators for their collegiality and professionalism.
%
%
\begin{acknowledgements}
This work was supported by the KESPRINT collaboration, an international consortium devoted to the characterisation and research of exoplanets discovered with space-based missions (www.kesprint.science).
JS, RK, MS, MK and PK would like to acknowledge support from MSMT grant LTT-20015.
JS and PK acknowledge a travel budget from ERASMUS+ grant 2020-1-CZ01-KA203-078200.
JS would like to acknowledge support from the Grant Agency of Charles University: GAUK No. 314421.
PC acknowledges the generous support from Deutsche Forschungsgemeinschaft (DFG) of the grant CH 2636/1-1. We are grateful for the generous support by Th\"uringer Ministerium f\"ur Wirtschaft, Wissenschaft und Digitale Gesellschaft.
K.W.F.L. was supported by Deutsche Forschungsgemeinschaft grants RA714/14-1 within the DFG Schwerpunkt SPP 1992, Exploring the Diversity of Extrasolar Planets.
NL was financially supported by the Ministerio de Economia y Competitividad and the Fondo Europeo de Desarrollo Regional (FEDER) under AYA2015-69350-C3-2-P\@.
EG is thankful for the generously supported by the by the Th\"uringer Ministerium f\"ur Wirtschaft, Wissenschaft und Digitale Gesellschaft and the staff of the Alfred-Jensch-Teleskop.
I.G., M.F.,  J. K., and C.M.P., gratefully acknowledge the support of the  Swedish National Space Agency (DNR 174/18, 2020-00104, 65/19).
This work was supported by JSPS KAKENHI grant number 20K14518.
R.L. acknowledges financial support from the Spanish Ministerio de Ciencia e Innovación, through project PID2019-109522GB-C52, and the Centre of Excellence "Severo Ochoa" award to the Instituto de Astrofísica de Andalucía (SEV-2017-0709).
LMS gratefully acknowledges financial support from the CRT foundation under Grant No. 2018.2323 ``Gaseous or rocky? Unveiling the nature of small worlds".
This work has been carried out within the framework of the NCCR PlanetS supported by the Swiss National Science Foundation.

We acknowledge the use of public TESS data from pipelines at the TESS Science Office and at the TESS Science Processing Operations Center. 
Resources supporting this work were provided by the NASA High-End Computing (HEC) Program through the NASA Advanced Supercomputing (NAS) Division at Ames Research Center for the production of the SPOC data products.
This paper includes data collected with the TESS mission, obtained from the MAST data archive at the Space Telescope Science Institute (STScI). Funding for the TESS mission is provided by the NASA Explorer Program. STScI is operated by the Association of Universities for Research in Astronomy, Inc., under NASA contract NAS 5–26555.
This work has made use of data from the European Space Agency (ESA) mission {\it Gaia} (\url{https://www.cosmos.esa.int/gaia}), processed by the {\it Gaia} Data Processing and Analysis Consortium (DPAC, \url{https://www.cosmos.esa.int/web/gaia/dpac/consortium}). Funding for the DPAC
has been provided by national institutions, in particular the institutions participating in the {\it Gaia} Multilateral Agreement.
This publication makes use of VOSA, developed under the Spanish Virtual Observatory project supported by the Spanish MINECO through grant AyA2017-84089. VOSA has been partially updated by using funding from the European Union's Horizon 2020 Research and Innovation Programme, under Grant Agreement nº 776403 (EXOPLANETS-A).
This research has made use of the SIMBAD database, operated at CDS, Strasbourg, France.
\end{acknowledgements}

\bibliographystyle{aa}
\bibliography{astro_citations}
\onecolumn
\appendix
\renewcommand\thefigure{\thesection.\arabic{figure}}    
\section{Additional material}

\begin{figure*}[ht!]
\centering
\includegraphics[width=1.0\textwidth]{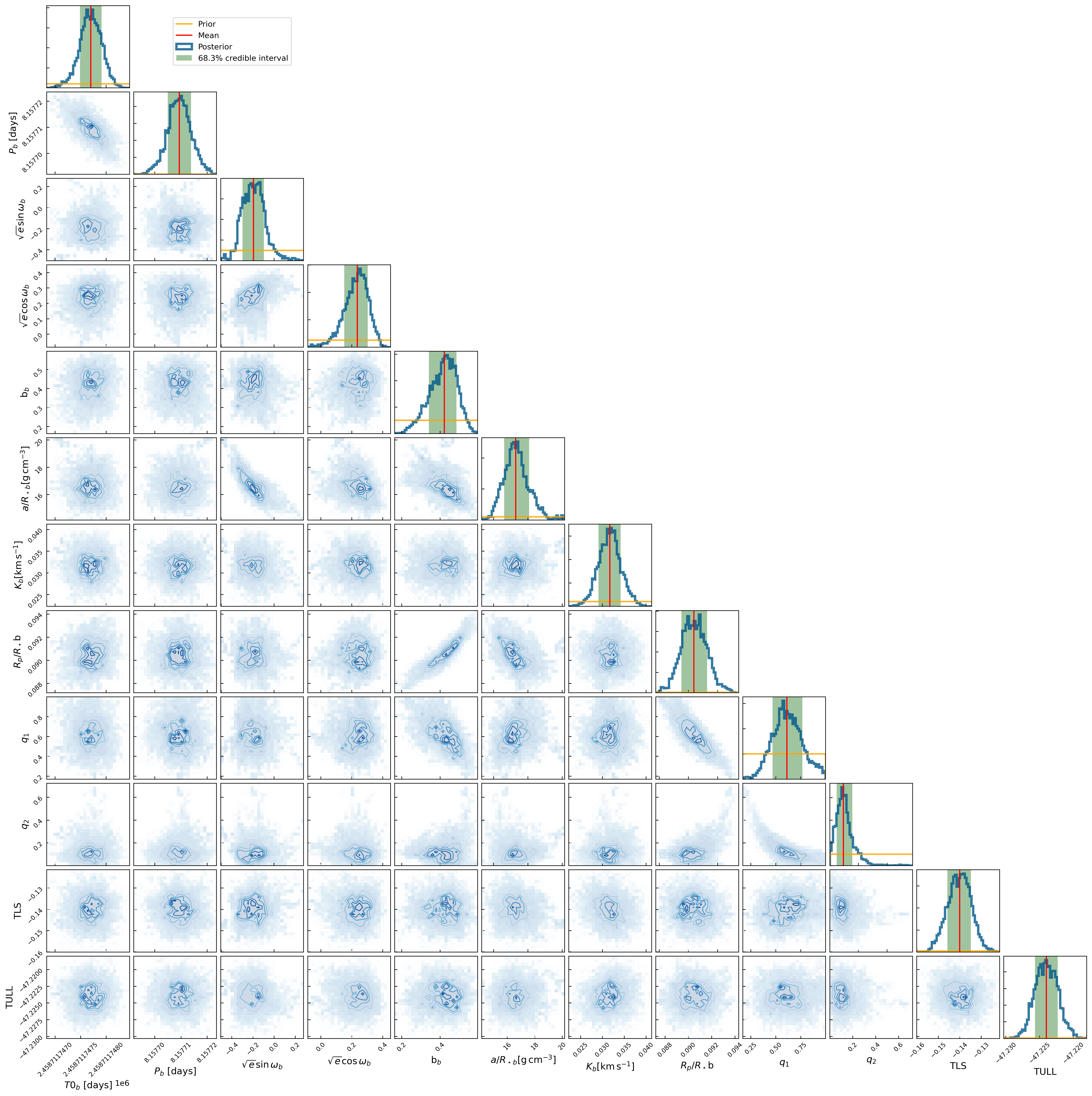}
\caption{The correlations between the free parameters from the MCMC analysis using the {\tt pyaneti} code. At the end of each row is shown the derived posterior probability distribution.} \label{fig:correlation}
\end{figure*}

\begin{table*}
 \centering
 \caption[]{Median values and 68\% confidence intervals for parameters from the {\tt pyaneti} analysis
 }
 \label{table:planet_par_pyan}
	\begin{tabular}{lcccr} 
		\hline 
		\hline
		Parameter & Unit & Value \\
		\hline
		\multicolumn{3}{c}{Input stellar parameters} \\
		\hline
        $M_\star$\dotfill & ($M_{\odot}$)\dotfill &  $ 0.96 \pm 0.04 $ \\
        $R_\star$\dotfill & ($R_{\odot}$)\dotfill & $0.92 \pm 0.06 $  \\
        $\rm T_{eff}$\dotfill & ($\mathrm{K}$)\dotfill & $ 5300 \pm 100 $ \\
        $v_{\rm rot} \sin{i_\star}\dotfill $ & $(km\,s^{-1}$)\dotfill & $ 4.1 \pm 1.0 $ \\
        J mag\dotfill & (mag)\dotfill & $ 9.40 \pm 0.02 $ \\
        \hline
        \multicolumn{3}{c}{Fitted parameters} \\
        \hline
        $T_0$\dotfill & (\bjdtdb)\dotfill &  $2458711.7477 _{ - 0.00021 } ^ { + 0.00021 }$ \\ 
        $P$\dotfill & (days)\dotfill &  $8.1577094 _{ - 4.4e-06 } ^ { + 4.5e-06 }$ \\ 
        $\sqrt{e} sin w$\dotfill & ()\dotfill & $-0.195 _{ - 0.104 } ^ { + 0.1 }$ \\ 
        $\sqrt{e} cos w$\dotfill & ()\dotfill & $0.236 _{ - 0.084 } ^ { + 0.067 }$ \\
        b\dotfill & ()\dotfill & $0.423 _{ - 0.081 } ^ { + 0.063 }$ \\ 
        a/$R_\star$\dotfill & ()\dotfill & $16.62 _{ - 0.84 } ^ { + 0.97 }$ \\ 
        $R_p/R_\star$\dotfill & ()\dotfill & $0.091 _{ - 0.001 } ^ { + 0.001 }$ \\ 
        K\dotfill & $(m\,s^{-1})$\dotfill & $31.7 _{ - 2.6 } ^ { + 2.5 }$ \\
        \hline
        \multicolumn{3}{c}{Derived parameters} \\
        \hline
        $M_p$\dotfill & ($M_{\oplus}$)\dotfill & $96.4 _{ - 8.3 } ^ { + 8.2 }$ \\ 
        $R_p$\dotfill & ($R_{\oplus}$)\dotfill & $9.1 _{ - 0.6 } ^ { + 0.6 }$ \\ 
        e\dotfill & ()\dotfill & $0.105 _{ - 0.039 } ^ { + 0.04 }$ \\ 
        $\omega$\dotfill & (deg)\dotfill & $-40.6 _{ - 19.5 } ^ { + 20.7 }$ \\ 
        i\dotfill & (deg)\dotfill & $88.63 _{ - 0.3 } ^ { + 0.32 }$ \\ 
        a\dotfill & (AU)\dotfill & $0.0711 _{ - 0.0058 } ^ { + 0.0063 }$ \\ 
        depth\dotfill & (ppm)\dotfill & $8186.0 _{ - 192.0 } ^ { + 205.0 }$ \\ 
        RM\dotfill & ($m\,s^{-1}$)\dotfill & $30.52 _{ - 7.48 } ^ { + 7.35 }$ \\ 
        Received irradiance\dotfill & ($F_{\oplus}$)\dotfill & $118.8 _{ - 15.1 } ^ { + 16.2 }$ \\ Transmission spectroscopy metric$^1$\dotfill & ()\dotfill & $127.7 _{ - 14.9 } ^ { + 17.1 }$ \\ 
        $\rho_\star$$^2$\dotfill & $(g\,cm^{-3})$\dotfill & $1.31 _{ - 0.19 } ^ { + 0.24 }$ \\ 
        $\rho_\star$$^3$\dotfill &$(g\,cm^{-3})$\dotfill & $1.74 _{ - 0.31 } ^ { + 0.39 }$ \\ 
        $T_{eq}$$^4$\dotfill & (K)\dotfill & $918.9 _{ - 30.7 } ^ { + 29.9 }$ \\ 
        $T_{tot}$\dotfill & (hours)\dotfill & $4.001 _{ - 0.024 } ^ { + 0.025 }$ \\ 
        $T_{full}$\dotfill & (hours)\dotfill & $3.206 _{ - 0.058 } ^ { + 0.049 }$ \\ 
        $\rho_p$\dotfill & $(g\,cm^{-3}$\dotfill & $0.71 _{ - 0.13 } ^ { + 0.17 }$ \\ 
        $g_p$$^5$\dotfill & $(cm\,s^{-2}$)\dotfill & $951.0 _{ - 135.0 } ^ { + 144.0 }$ \\ 
        $g_p$$^6$\dotfill & $(cm\,s^{-2}$)\dotfill & $1147.0 _{ - 165.0 } ^ { + 198.0 }$ \\ 
        Jeans Escape Parameter$^7$\dotfill & ()\dotfill & $87.54 _{ - 9.62 } ^ { + 10.79 }$ \\ 
        \hline
        \multicolumn{3}{c}{Other parameters} \\
        \hline
        $q_1$\dotfill & ()\dotfill & $0.61 _{ - 0.14 } ^ { + 0.15 }$ \\ 
        $q_2$\dotfill & ()\dotfill & $0.119 _{ - 0.059 } ^ { + 0.077 }$ \\ 
        $u_1$\dotfill & ()\dotfill & $0.191 _{ - 0.09 } ^ { + 0.088 }$ \\ 
        $u_2$\dotfill & ()\dotfill & $0.6 _{ - 0.18 } ^ { + 0.15 }$ \\ 
        Sys. vel. TCES\dotfill & $(km\,s^{-1}$)\dotfill & $-0.14 _{ - 0.0058 } ^ { + 0.0052 }$ \\ 
        Sys. vel. Tull\dotfill & $(km\,s^{-1}$)\dotfill & $-47.224 _{ - 0.0017 } ^ { + 0.0017 }$ \\
        \hline
	\end{tabular}
\smallskip\\
Comments: 1 - based on the equation from \citet{Kempton18}; 2 - based on the equation from \citet{Winn10}, 3 - from stellar parameters 4 - based on the equation from \citet{Kempton18}; 5 - from K \& $R_p/R_\star$, 6 - from planetary parameters; 7 - based on the equation from \citet{Fossati17} \\
\end{table*}

\onecolumn

\begin{longtable}{lccc}
\caption{Relative radial velocities of TOI-1268 from Tautenburg and McDonald.} \label{tab:long} \\

\hline \multicolumn{1}{c}{\textbf{Date (BJD)}} & \multicolumn{1}{c}{\textbf{RV (km/s)}} & \multicolumn{1}{c}{\textbf{$\sigma_{RV}$ (km/s)}} & 
\multicolumn{1}{c}{\textbf{Instrument}} \\ \hline 
\endfirsthead

\multicolumn{4}{c}%
{{\bfseries \tablename\ \thetable{} -- continued from previous page}} \\
\hline \multicolumn{1}{c}{\textbf{Date (BJD)}} & \multicolumn{1}{c}{\textbf{RV (km/s)}} & \multicolumn{1}{c}{\textbf{$\sigma_{RV}$ (km/s)}} & \multicolumn{1}{c}{\textbf{Instrument}} \\ \hline 
\endhead

\hline \multicolumn{4}{r}{{Continued on next page}} \\ 
\endfoot

\hline 
\endlastfoot

2459191.983808	& -47.22437	& 0.00856	& Tull \\
2459192.970784	& -47.24564	& 0.01418	& Tull \\
2459203.007863	& -47.23507	& 0.01196	& Tull \\
2459204.007530	& -47.23584	& 0.01461	& Tull \\
2459240.846169	& -47.19582	& 0.01313	& Tull \\
2459269.908577	& -47.22845	& 0.00749	& Tull \\
2459270.919200	& -47.22343	& 0.00765	& Tull \\
2459275.783607	& -47.22655	& 0.01299	& Tull \\
2459275.883947	& -47.26212	& 0.00641	& Tull \\
2459276.880298	& -47.26054	& 0.01158	& Tull \\
2459277.696787	& -47.22547	& 0.01169	& Tull \\
2459281.016337	& -47.17902	& 0.01052	& Tull \\
2459281.913238	& -47.20045	& 0.00487	& Tull \\
2459293.865327	& -47.24465	& 0.01072	& Tull \\
2459294.862017	& -47.27227	& 0.01130	& Tull \\
2459301.896369	& -47.26636	& 0.01316	& Tull \\
2459302.874042	& -47.25431	& 0.00832	& Tull \\
2459308.793593	& -47.24278	& 0.00949	& Tull \\
2459309.781640	& -47.24384	& 0.01045	& Tull \\
2459339.735803	& -47.22857	& 0.00886	& Tull \\
2459340.750300	& -47.23501	& 0.00534	& Tull \\
2459355.656311	& -47.20177	& 0.01085	& Tull \\
2459355.826286	& -47.20422	& 0.01092	& Tull \\
2459372.686595	& -47.22036	& 0.00788	& Tull \\
2459383.816171	& -47.25009	& 0.00897	& Tull \\
2459384.661582	& -47.21950	& 0.01028	& Tull \\
2459384.769060	& -47.21000	& 0.01148	& Tull \\
2459385.678300	& -47.21702	& 0.00790	& Tull \\
2459385.812858	& -47.21691	& 0.01250	& Tull \\
2459411.662985	& -47.19000	& 0.00956	& Tull \\
2459412.715212	& -47.20674	& 0.01116	& Tull \\
2459413.656790	& -47.21929	& 0.01094	& Tull \\
2458913.408718	& -0.11477	& 0.06461	& TCES \\
2458913.457617	& -0.16183	& 0.03256	& TCES \\
2458916.547478	& -0.21934	& 0.03111	& TCES \\
2458916.569227	& -0.18508	& 0.04616	& TCES \\
2458918.459735	& -0.26428	& 0.03768	& TCES \\
2458918.482571	& -0.28070	& 0.03562	& TCES \\
2458921.466170	&  0.03977	& 0.03811	& TCES \\
2458921.487907	& -0.00825	& 0.02823	& TCES \\
2458923.447120	& -0.10471	& 0.03872	& TCES \\
2458923.470291	& -0.04594	& 0.02784	& TCES \\
2459026.409863	& -0.07805	& 0.02322	& TCES \\
2459027.465916	& -0.09616	& 0.02724	& TCES \\
2459067.447038	& -0.14904	& 0.03380	& TCES \\
2459068.395396	& -0.10968	& 0.02112	& TCES \\
2459068.514666	& -0.08156	& 0.04332	& TCES \\
2459095.580327	& -0.12706	& 0.04659	& TCES \\
2459099.577269	& -0.22278	& 0.10554	& TCES \\
2459100.503684	& -0.12215	& 0.04659	& TCES \\
2459101.466692	& -0.06668	& 0.04494	& TCES \\
2459177.671762	& -0.14198	& 0.04193	& TCES \\
2459178.637930	& -0.19723	& 0.03451	& TCES \\
2459178.660246	& -0.24076	& 0.05164	& TCES \\
2459179.542871	& -0.20752	& 0.05887	& TCES \\
2459179.564597	& -0.17839	& 0.05570	& TCES \\
2459209.524158	& -0.27055	& 0.03726	& TCES \\
2459209.639666	& -0.14617	& 0.03737	& TCES \\
2459213.601844	& -0.12355	& 0.03448	& TCES \\
2459246.401565	& -0.07916	& 0.05877	& TCES \\
2459246.506298	& -0.07395	& 0.04963	& TCES \\
2459265.503996	& -0.13282	& 0.03105	& TCES \\
2459265.525721	& -0.01149	& 0.03959	& TCES \\
2459266.523684	& -0.19435	& 0.03207	& TCES \\
2459266.545409	& -0.15557	& 0.06025	& TCES \\
2459268.453833	& -0.18619	& 0.03689	& TCES \\
2459268.654701	& -0.13891	& 0.04656	& TCES \\
2459268.676506	& -0.13444	& 0.07028	& TCES \\
2459269.297408	& -0.18628	& 0.04416	& TCES \\
2459270.304122	& -0.25730	& 0.04798	& TCES \\
2459271.279054	& -0.21693	& 0.03867	& TCES \\
2459271.602628	& -0.24096	& 0.04051	& TCES \\
2459272.629216	& -0.18709	& 0.04589	& TCES \\
2459276.676413	& -0.14742	& 0.04496	& TCES \\
2459298.410037	& -0.11411	& 0.03411	& TCES \\
2459303.538520	& -0.15347	& 0.02992	& TCES \\
2459304.516020	& -0.07711	& 0.02653	& TCES \\
2459304.537744	& -0.05797	& 0.02515	& TCES \\
2459305.524122	& -0.11802	& 0.02205	& TCES \\
2459305.545847	& -0.07095	& 0.03662	& TCES \\
2459309.539517	& -0.09818	& 0.07137	& TCES \\
2459310.526783	& -0.12355	& 0.06796	& TCES \\
2459240.478590	& -0.13677	& 0.02861	& TCES \\
\hline

\end{longtable}

\end{document}